\documentclass[prapplied,superscriptaddress,reprint,amsmath,amssymb,aps,longbibliography]{revtex4-2}

\usepackage{graphicx}
\usepackage{float}
\usepackage[
pdfstartview=FitH,
bookmarks=true,
breaklinks=true,
colorlinks=true,
linkcolor=blue,anchorcolor=blue,
citecolor=blue,filecolor=blue,
menucolor=blue,urlcolor=blue]{hyperref}

\newcommand{\set}[1]{\mathbb{#1}}
\newcommand{\map}[1]{\mathcal{#1}}

\begin{document}

\title{Accelerating the assembly of defect-free atomic arrays with maximum parallelisms}

\author{Shuai Wang}
\thanks{These authors contribute equally to this work.}
\author{Wenjun Zhang}
\thanks{These authors contribute equally to this work.}
\author{Tao Zhang}
\thanks{These authors contribute equally to this work.}
\author{Shuyao Mei}
\author{Yuqing Wang}
\affiliation{Department of Physics and State Key Laboratory of Low Dimensional Quantum Physics, Tsinghua University, Beijing, 100084, China}
\author{Jiazhong Hu}
\author{Wenlan Chen}
\email{cwlaser@ultracold.cn}
\affiliation{Department of Physics and State Key Laboratory of Low Dimensional Quantum Physics, Tsinghua University, Beijing, 100084, China}
\affiliation{Frontier Science Center for Quantum Information, Beijing, 100084, China}
\affiliation{Collaborative Innovation Center of Quantum Matter, Beijing, 100084, China}

\date{\today}

\begin{abstract}
Defect-free atomic arrays have been demonstrated as a scalable and fully-controllable platform for quantum simulations and quantum computations. To push the qubit size limit of this platform further, we design an integrated measurement and feedback system, based on field programmable gate array (FPGA), to quickly assemble two-dimensional defect-free atomic array using maximum parallelisms.
The total time cost of the rearrangement is first reduced by processing atom detection, atomic occupation analysis, rearrangement strategy formulation, and acousto-optic deflectors (AOD) driving signal generation in parallel in time. Then, by simultaneously moving multiple atoms in the same row (column), we save rearrangement time by parallelism in space. To best utilize these parallelisms, we propose a new algorithm named Tetris algorithm to reassemble atoms to arbitrary target array geometry from two-dimensional stochastically loaded atomic arrays. For an $L \times L$ target array geometry, the number of moves scales as $L$, and the total rearrangement time scales at most as $L^2$. 
Although in this work we do not test on actual atoms, we simulate the performance of our FPGA system experimentally with all components integrated except for the atoms.
We present the overall performance for different target geometries, and demonstrate a dramatic boost in rearrangement time cost and the potential to scale up defect-free atomic array to 1000 atoms in room-temperature platform and 10000 atoms in cryogenic environment.

\end{abstract}

\maketitle

\section{Introduction}

\begin{figure*}[htbp]
    \centering
    \includegraphics[width=0.78\linewidth]{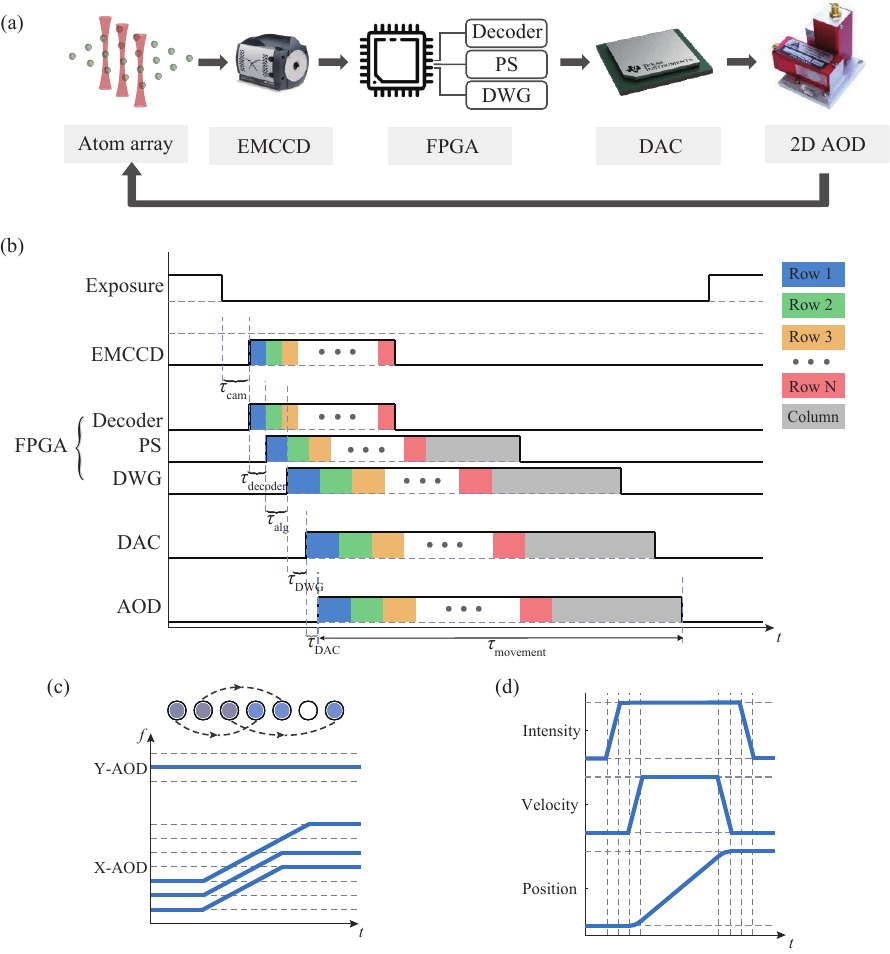}
    \caption{ 
        (a) The rearrangement control flow. 
        (b) A typical time sequence of our system. Each row shows the operation periods for one module or device, where a high-level signal indicates that the module is in operation. The blue, green, yellow, and red blocks indicate the periods of processing the first, second, third, and the last row of atoms, respectively. By organizing these modules to work in parallel, the total time cost $\tau$ is significantly reduced.
        (c) The spectra of the AOD driving signals to simultaneous rearrange atoms in a row. The Y-AOD, which controls the tweezer position in vertical direction, requires only a monochromatic driving signal, while the X-AOD requires a multi-tone waveform. 
        (d) Improving the quality of mobile tweezer controls to avoid atom loss or heating during the rearrangement. When a tweezer begins to move, its trap depth is ramped up to smoothly transfer the atom to the mobile tweezers from static traps. Then the tweezer accelerates to the maximum velocity with a fixed acceleration. This process is reversed when approaching to the final destination.
    } \label{fig:principal_of_operation}
\end{figure*}

Individual neutral atoms trapped in optical tweezers is a fast evolving toolbox featuring unprecedented high controllability for quantum systems~\cite{kaufman_ni_2021,browaeys2020many,doi:10.1126/science.aav9105,bernien2017probing,blackmore2018ultracold,21-256-Lukin,16-51-Lukin,20Enhanced-assembly,22-289-Lukin,graham2022multi,singh2022mid,kasper2021universal,preskill2018quantum,wilson2022trapping,2022-schymik2022situ}. It has demonstrated its scalability and programmability in many applications. Quantum simulations on numerous topics have benefited from its development, ranging from exotic quantum matters to fundamental theories underlying the universe~\cite{browaeys2020many,21-256-Lukin,doi:10.1126/science.aav9105,blackmore2018ultracold,bernien2017probing}. It is also one of the most promising platforms for quantum computation in the noisy intermediate-scale quantum (NISQ) era~\cite{22-289-Lukin,singh2022mid, graham2022multi,kasper2021universal,preskill2018quantum}. Most applications mentioned above require each neutral atom to be arranged into arrays of various desired geometries without defects. However, the process of loading single atoms into optical tweezers is stochastic~\cite{PhysRevLett.89.023005,schlosser2001sub}. In order to obtain a defect-free atomic array, it is common to rearrange the initially loaded atoms to the target positions using a set of mobile tweezers. Using such rearrangement method, defect-free atomic arrays of various geometries with hundreds of atoms have been demonstrated~\cite{21-256-Lukin,20Enhanced-assembly,16-51-Lukin}. 

In order to further scale up the size of defect-free atomic arrays, a key challenge lies at the tension between the decreasing lifetime of the atomic ensemble and the increasing time cost of rearranging the atoms~\cite{PhysRevX.9.011057}. Two major strategies have been developed to overcome this challenge. One is to extend the vacuum limited lifetime. For example, by placing the system in a cryogenic environment, a single-atom lifetime of 6000 seconds and defect-free arrays with 324 atoms are reported recently~\cite{2022-schymik2022situ,PhysRevApplied-6000s}. The other strategy is to reduce the rearrangement time cost by designing faster rearrangement protocols. Based on single mobile optical tweezers produced by AODs, a variety of rearrangement algorithms have been demonstrated, such as the shortest-move heuristic algorithm~\cite{19MLA,PhysRevResearch.3.023008}, intuitive algorithm~\cite{16Browaeys,20Enhanced-assembly}, and Hungarian matching algorithm~\cite{graham2022multi}. 

In recent works, multi-mobile-tweezer algorithms have also been demonstrated~\cite{tian2022parallel,21-256-Lukin} and show their capability to assemble larger defect-free atomic arrays. To further decrease the time cost of atom rearrangement, and increase the size of defect-free atomic arrays, we need to lift two hardware limitations. One is that people usually need to wait until a full atomic fluorescence image has been read out from an electron-multiplying CCD (EMCCD) camera to determine atomic occupations, despite the fact that in some cases only part of the atomic image is needed to make corresponding rearrangement strategies. 
The other is that, with the widely-used arbitrary waveform generators (AWGs), the maximum number of simultaneous controlled mobile tweezers is limited by the slow speed of the calculation and streaming of the control signals for the mobile tweezers.

In this work, we lift these limitations by designing an FPGA-based integrated measurement and feedback system featuring maximum parallelisms for the rearrangement. We integrate the abilities of low-latency atomic occupation analysis, rearrangement strategy formulation, and real-time multi-tweezer control in this system, so that these processes work in parallel to save in-sequence waiting time. We develop a new rearrangement algorithm, named Tetris algorithm, which is tailored for the parallel hardware structure. These developments significantly reduce the rearrangement time cost, and thus pave the way to rearrangement and assembly of large-scale defect-free atomic arrays.

This manuscript is organized as follows. 
We begin by introducing the principle of operations in Sec.~\ref{sec:overview}, where we present the method of decoding atomic occupation data and generating control signals for mobile tweezers in parallel, and then the Tetris algorithm. The details of the hardware implementation are given in Sec.~\ref{sec:control_system}, as well as the experimental characterizations of the rearrangement time complexity. Finally, we give the conclusion and outlook in Sec.~\ref{sec:conclusion}.

\section{Principle of operation}\label{sec:overview}

In order to assemble a large-scale defect-free atomic array, the typical experimental scheme is to first stochastically load the atoms into an array of static optical tweezers and use fluorescence imaging to determine the positions of defects. Then another set of mobile tweezers are used to move the loaded atoms into target positions so that the atomic array is rearranged into desired geometries. Here, we focus on parallelizing the processes of image readout, rearrangement strategy making, and atom movement to minimize the rearrangement time cost. We introduce the principle of operation of the system in this section and leave technical details to Sec.~\ref{sec:control_system}.

\subsection{Overview of the rearrangement control flow}

Our rearrangement control flow is shown in Fig.~\ref{fig:principal_of_operation}(a). The system consists of an EMCCD, an FPGA, a digital-to-analog converter (DAC), and a two-dimensional AOD. Instead of using a computer and an AWG, we integrate fluorescence image processing, atomic occupation analysis, rearrangement strategy formulation, and the generation of the control signals for AODs as digital circuits on a single FPGA chip. This integration not only enables fast communications and fully parallel operation of all the procedures, but also provides high-quality and scalable real-time generation of the multi-tone RF signals, which allows us to move almost arbitrarily many tweezers simultaneously. 

In one experiment cycle, the atoms are first loaded stochastically into a static array of tweezers. Then a fluorescence image is taken by the EMCCD to detect atomic occupation.
It is important that the image data is transferred to the image decoder module on the FPGA pixel by pixel, so that one can begin to rearrange one row of atoms as soon as their occupation data is extracted. That is, when the pixel data of the next row is still being transferred to the FPGA, the on-FPGA soft-core processor system (PS) already begins to run the specially designed Tetris algorithm (see Sec.~\ref{subsec:algorithm}) to determine the rearrangement strategy of this row and to instruct a digital waveform generator (DWG) module to produce the corresponding multi-tone driving waveform. The generated waveforms are then directed to DACs and sent to AODs to control the optical tweezers.

To demonstrate how this system maximizes parallelism, we show a typical time sequence in Fig.~\ref{fig:principal_of_operation}(b). We program the modules on the FPGA to work in a pipelined fashion. Once the atom imaging exposure has completed, the EMCCD begins to send the photon-counting data pixel by pixel after a short constant internal delay $\tau_{\rm cam}$. The image decoder receives this data almost immediately ($<10 ~\rm ns$) and starts to analyze whether each static tweezer is occupied with an atom. Such decoding time of atom occupation information for one row of tweezers is $\tau_{\rm decoder}$. Once the first row of tweezers (blue in Fig.~\ref{fig:principal_of_operation}(b)), instead of the full image, have been analyzed, the decoder sends the atom occupation data to the PS which uses the Tetris algorithm to determine a rearrangement strategy for this row with time cost $\tau_{\rm alg}$. Then, the DWG and the DAC start to work successively, with latency $\tau_{\rm DWG}$ and $\tau_{\rm DAC}$, respectively. 
It is important to note that when the PS is running the algorithm, the decoder is simultaneously analyzing the image data of the second row (green in Fig.~\ref{fig:principal_of_operation}(b)). Different modules work simultaneously, instead of waiting until the previous one finishes processing, and the atoms in one row are moving to their target positions while the image data of another row in EMCCD is being read out.

In summary, the total time cost $\tau$ is significantly reduced and is determined by a starting overhead and the atom movement time $\tau_{\text{movement}}$ (Fig.~\ref{fig:principal_of_operation}(b)). The starting overhead includes the internal processing time of the EMCCD $\tau_{\text{cam}}$ and the processing time of the first row of atoms, including the analysis time of the decoder $\tau_{\text{decoder}}$, the time cost of running the Tetris algorithm $\tau_{\text{alg}}$, the DWG programming time $\tau_{\text{DWG}}$, and the DAC conversion delay $\tau_{\text{DAC}}$. Therefore, the total time cost is limited only by the most time-consuming module, instead of the summation of all modules. In the large-scale limit, it is dominated by $\tau_{\text{movement}}$.

\subsection{Simultaneously rearranging atoms in one row}\label{subsection:parallel_tweezer}
 
We further reduce the overall time cost by moving all the atoms in a row simultaneously. Even in the worst case scenario, the one-row rearrangement time only scales as $L$ for an $L\times L$ array, compared with $\sim L^2$ in the case that only one tweezer is moved at a time. This requires a multi-tone waveform $S(t) = \sum_{i=1}^K A_i \cos (\varphi_i(t))$ generation to control $K$ simultaneous mobile tweezers.
In Fig.~\ref{fig:principal_of_operation}(c), we show a typical movement of mobile tweezers and the frequency spectra of the control signals for X- and Y-AODs, where the X-AOD (or Y-AOD) controls the horizontal (vertical) positions of the mobile tweezers. The Y-AOD control signal maintains at a constant frequency, while the X-AOD control signal features multiple changing frequencies. 
Here, we use multiple direct digital synthesizers (DDSs) in the DWG (see Sec.~\ref{sec:control_system} for details) to generate the desired multi-tone waveform in real time. 
This minimizes the internal processing time and data storage at the cost of FPGA on-chip resources. 
Given capabilities of current commercial FPGAs, real-time signal generation to control more than $400$ mobile tweezers is achievable.

The quality of mobile tweezers can also benefit from this method. The DWG structure intrinsically guarantees the phase-continuity of the driving signals and avoids potential kicks to the atoms. 
Additional improvements can also be applied to avoid atom loss or heating during the tweezer moving. As demonstrated in Fig.~\ref{fig:principal_of_operation}(d), for a higher speed and success rate, strategies such as turning on and off the moving tweezers smoothly~\cite{16Browaeys}, accelerating or decelerating atoms smoothly at the beginning or the end of the tweezer moving~\cite{16-51-Lukin}, and setting a maximum moving speed~\cite{20Enhanced-assembly} are all realized in our system. Therefore, the ability of multi-tweezer manipulation with maximum flexibility and low latency is accessible through our system.

\begin{figure*}[htbp]
    \centering
    \includegraphics[width=\linewidth]{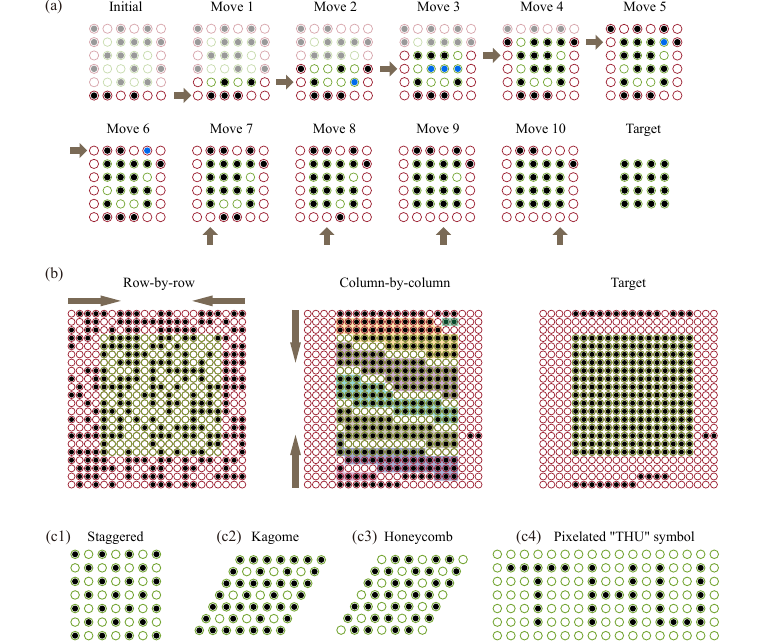}
    \caption{ 
        Illustration of the Tetris algorithm. 
        (a) The rearrangement strategy of a $4\times 4$ target square lattice. Moves 1--6 are row-by-row rearrangements. In each move, we first horizontally align atoms with the defects in the previous move (marked blue). Then the atoms left are assigned to the leftmost sites in the target geometry. This does not require knowledge of the next rows of atoms (marked transparent). The row rearrangement ensures that each column has enough atoms required by the target array. Moves 7--10 are column-by-column compression where atoms are moved to the final target positions.
        (b) The rearrangement of a larger array. The row-by-row rearrangements organize the atoms into blocks (Tetriminoes). The column-by-column compression arrange atoms into target geometry (Tetrimino eliminations).
        (c) Examples of arbitrary target geometries achieved with Tetris algorithm, by rotating the AODs or pixelating the desired pattern: staggered (c1), Kagome (c2), honeycomb (c3) lattices, and the pixelated ``THU'' symbol (c4).
    } \label{fig:tetris}
\end{figure*}

In comparison, in a computer-AWG structure, though these features can in principle also be realized, its response time is slow. And the number of simultaneous mobile tweezers is limited if one pre-calculates and stores all the possible needed waveforms in advance.
In such systems, the computer needs to calculate the waveform at each time step and transfer this data to the AWG. This process is carried out by a software on the computer instead of by direct hardware implementation.
Therefore, on the one hand, if the waveform is calculated in real time, both the calculation and data transfer costs significant time. On the other hand, if one calculates all the possible needed waveforms in advance and stores them in the buffer of the AWG, the storage resource consumption grows exponentially for exhausting all combinations of the movement as the total number of mobile tweezers increases.

\subsection{The Tetris algorithm} \label{subsec:algorithm}

To best utilize the parallelism of our system, we develop the Tetris algorithm to give the rearrangement strategy, inspired by the game Tetris where players need to adjust Tetriminoes (i.e., blocks) to fulfill and eliminate one complete row. 
In the following, we explain this algorithm first in a simple example, and then in strict mathematical description.

In the example, we target at a defect-free compact $4\times 4$ square lattice. As shown in Fig.~\ref{fig:tetris}(a) with the target geometry marked in green, we start with a stochastically loaded $6\times 6$ array with filling factor of 50\%~\cite{schlosser2001sub}. Initial array with larger size could be used in real experiments to guarantee that enough atoms are loaded. 

In the first move, we arrange the first row of atoms to horizontally align with the leftmost target sites, i.e., sites in the target geometry, and record the unfilled sites as defects. 
These atoms will be vertically moved to the first row of target sites in future steps.
At this point, no information of other rows of atoms is needed (thus marked transparent in Fig.~\ref{fig:tetris}(a)). In the second move, we first arrange the atoms to horizontally align with the defects left by the first move (named borrowed atoms and marked blue). Then the atoms left are again arranged to the leftmost target sites and the defects left are recorded. 
The borrowed atoms are still marked as defects, because they will eventually fill the defects left by the first move and be moved to the first row of target sites in future steps, instead of the second row.
Then in moves 3--6, we repeat the above procedures by first moving the atoms to horizontally align with the defects left by the previous move, and then arranging other atoms to align with the leftmost target sites. It is worth emphasizing that, although we describe the moving of borrowed atoms and ordinary atoms in sequence, all of them in the same row are actually moved simultaneously. 

Moves 1--6 are called Tetrimino constructions, since it is clear that after move 6, the atoms are arranged as pieces of connected blocks (more visible in a larger array, see the middle panel of Fig.~\ref{fig:tetris}(b)). 
At this point we check if there are enough atoms in each column. We proceed only when this is satisfied. 
Otherwise, we discard this attempt and start over.
The next moves are named Tetrimino elimination. In move 7--10, we compress the columns in the target area in the vertical direction to form defect-free columns, and obtain the target defect-free $4\times 4$ atomic array.

Now we give the strict mathematical description of the Tetris algorithm to rearrange target arrays with arbitrary geometries. We start with an $L \times W$ reservoir tweezer array. The target geometry is described by $L$ ordered sets, $\mathbb{R}_1, \mathbb{R}_2, \dots, \mathbb{R}_L$, where $\mathbb{R}_i = \left[ r_{i,1}, r_{i,2}, \dots, r_{i,N_i} \right]$, with $N_i<W$, records the row indices of the target atomic positions in the $i$-th column. The total atom number in the target array is $N = \sum_{i=1}^L N_i$. As mentioned in the example above, the algorithm takes three steps.

\paragraph{Row-by-row rearrangements (Tetrimino construction).}
Each row-by-row movement rearranges the initially loaded atoms in one row of reservoir tweezers to construct the building blocks for the target geometry. Mathematically, this requires us to assign each element in each ordered set $\set{R}_i$ with a loaded atom. Take the notation that the $k$-th row has $n^{(k)}$ initially loaded atoms and their column indices are recorded in an ordered set $\set{C}^{(k)}\equiv \left[c_{i}^{(k)}\right]$ which is in ascending order. For each column we use an ordered set $\set{R}_i^{(k)}$ to record the unassigned elements in $\set{R}_i$ before the move $k$. Naturally, $\set{R}_i^{(1)}=\set{R}_i$.

The rearrangement strategy for this row is given by an ordered set $\set{T}^{(k)}\equiv \left[t_{i}^{(k)}\right]$ describing the column indices which loaded atoms should be moved to, and a strategy mapping $\map{S} : \set{C}^{(k)} \mapsto \set{T}^{(k)}$ describing the specific movements. To determine $\set{T}^{(k)}$, we first look for the smallest element in each ordered set $\set{R}_i^{(k)}$ and record them in an ordered set $\set{M}=\left[ m_i \right]$ with $m_i=\min \set{R}_i^{(k)}$. Then we sort $\set{M}$ in ascending order, $\set{M}^{\prime} \equiv \left[ m_i^{\prime} \right] = \text{Sort}\left(\set{M}\right)= [m_{p_i}]$, where $\map{P} \equiv \left[p_i\right]$ is the sorting permutation. Then $\set{T}^{(k)}$ is given by the first $n^{(k)}$ elements in the permutation, i.e., $t_{i}^{(k)}=p_i$ for $i\le n^{(k)}$. Then we sort $\set{T}^{(k)}$ with permutation $Q\equiv \left[q_i\right]$ so that $\left[t_{q_i}^{(k)}\right]$ is in ascending order. This sorting avoids the atomic collisions during the rearrangement. The strategy $\map{S}$ is then naturally given by $\map{S}\left(c_{i}^{(k)}\right) = t_{q_i}^{(k)}$.
Finally, the set $\set{R}_i^{(k+1)}$ for the next row is determined by
$$
\set{R}_i^{(k+1)}=
\begin{cases}
	\set{R}_i^{(k)}\setminus\left\{m_i\right\} &\,, i\in \set{T}^{(k)} \\
	\set{R}_i^{(k)} &\,, i\notin \set{T}^{(k)}
\end{cases}
$$

\paragraph{Examination and post selection.}
After the row-by-row rearrangements, we check if $\set{R}_i^{(k+1)}$ is empty for all $i$. If this is not satisfied, unfortunately some columns have insufficient atoms. In this case, this experimental cycle is abandoned.
 
\paragraph{Column-by-column compression (Tetrimino elimination).}
Once previous steps succeed, only a straightforward compression for each column is needed to move the atoms to the desired positions. After this, we obtain the target geometry.
 
This procedure does not depend on the specific forms of the sets $\set{R}_i$. By choosing appropriate $\set{R}_i$, arbitrary geometries can be achieved conveniently. Note that despite the notation, the rows and columns are not necessarily perpendicular to each other. Therefore, geometries like honeycomb lattices or Kagome lattices can be easily realized by simply rotating the AODs to an angle of $60^{\circ}$ to each other. Some examples of different geometries are shown in Fig.~\ref{fig:tetris}(c).
 Following this logic, arbitrary array geometries can be implemented, as long as they are pixelated. This algorithm fully exploits the hardware advantages of parallelism, since the only needed data for rearranging a row of atoms is the occupation of current and previous rows.

\begin{figure}[htbp]
    \centering
    \includegraphics[width=\linewidth]{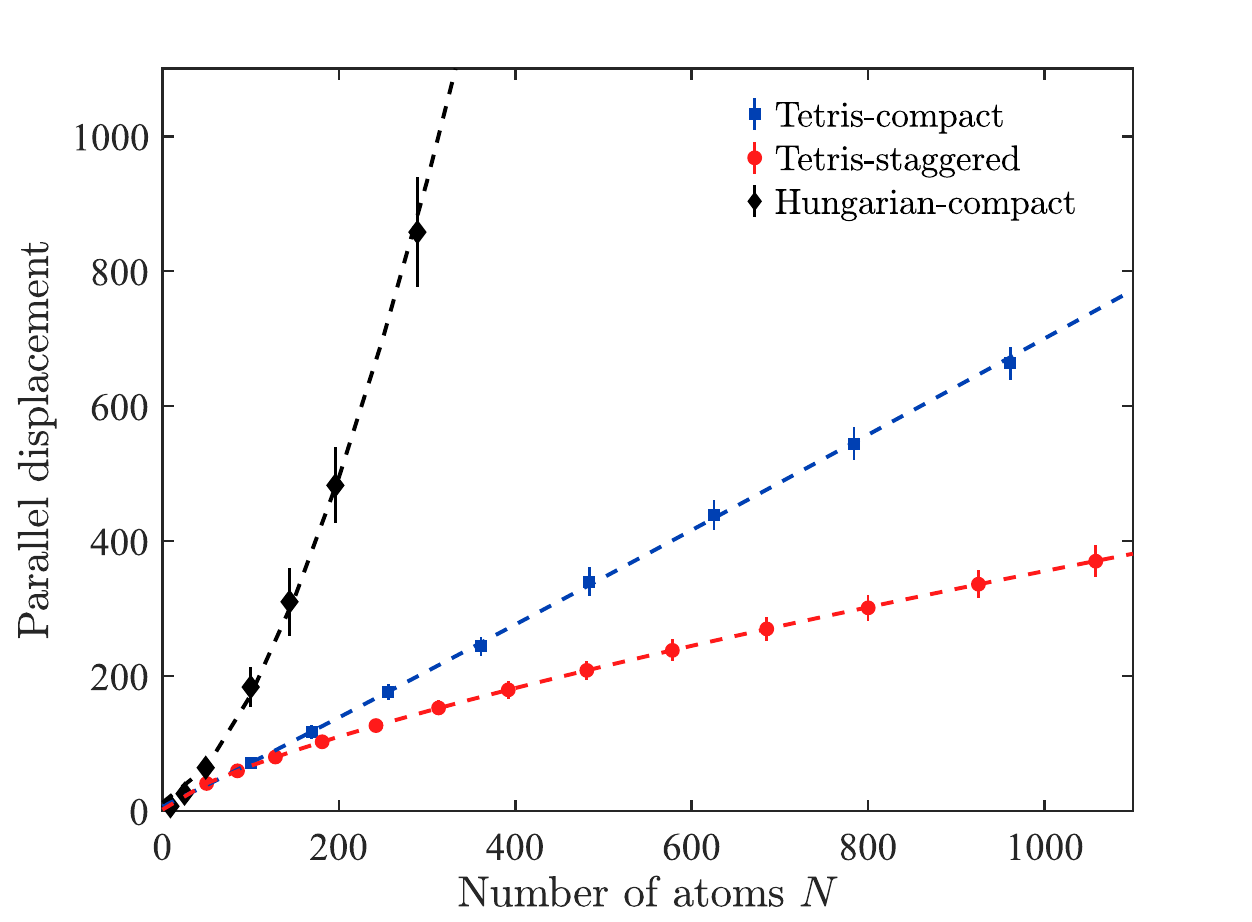}
    \caption{ 
        Monte Carlo simulation of the rearrangement complexity: the number of parallel displacements as a function of total atom number $N$ for different algorithms and target geometries. Blue and red dots represent compact and staggered geometries based on Tetris algorithm, while black dots represent compact geometry based on Hungarian algorithm for comparison. Each point is averaged over 10000 simulations with the error bar showing the standard deviations. The dashed lines show the power-law fitting with scaling factor of $N^{1.6(1)}$ for compact geometry with Hungarian algorithm, $N^{1.03(7)}$ for compact geometry with Tetris algorithm, and $N^{0.736(6)}$ for staggered geometry with Tetris algorithm when the initial loading efficiency is 50\%.
    }\label{fig:complexity}
\end{figure}

We verify the performance of the Tetris algorithm by Monte Carlo simulations.
In the simulations, the initial loading efficiency is set to $50\%$~\cite{schlosser2001sub}. 
For an $L \times L$ compact array, we start with a $\lceil (\sqrt{2}L+1)\rceil \times \lceil (\sqrt{2}L+1)\rceil$ array to ensure sufficient number of atoms are initially loaded in the vast majority of cases. Here $\lceil \cdot \rceil$ denotes the upper rounding integer.
And for an $L \times L$ staggered pattern, we start with an $(L+1) \times (L+1)$ array. 
In all of our simulations, only 0.18\% cases fail due to insufficient initial number of atoms.

As demonstrated in Ref.~\cite{21-256-Lukin}, the time cost in rearrangement is dominated by the time of atom travelling between different sites. This time is directly related to the number of parallel displacement (i.e., the sum of the largest atomic displacement in each move), which is shown for different algorithms and geometries in Fig.~\ref{fig:complexity}. 
Here, one atomic displacement is defined as the movement between two adjacent sites.
For an $L \times L$ configuration, there are approximately $L$ row moves and $L$ column moves needed, so the number of moves scales as $L$. In each move, the largest displacement is at most proportional to $L$. 
Therefore, the number of parallel displacements scales as $L^2\propto N$ for the worst-case scenario, in agreement with the simulation results.
As a comparison, the Hungarian algorithm, which is the strict optimal solution to the rearrangement problem without parallel tweezer movements, needs $\sim L^3$ atomic displacements.
These results indicate that the Tetris algorithm enables faster rearrangement than the solutions based on one-by-one movements. 

We also notice that better scaling could be achieved with the Tetris algorithm in some special cases. For example, the simulations show that for staggered pattern with 50\% initial loading efficiency, the number of parallel displacements scales as $N^{0.736(6)}$.
 
\section{Hardware structure design and characterization}\label{sec:control_system}
 
In this section, we describe the design of hardware structure of the control system in details and provide experimental characterizations of the time complexities. Most parts of the processing unit are integrated on an FPGA, which consists of three relatively independent modules. That is, an image decoder for row-by-row readout, a PS for rearrangement strategy making using the Tetris algorithm, and a DWG for the scalable driving signal generation of multiple mobile tweezers.
 
\begin{figure}[htbp]
    \centering
    \includegraphics[width=\linewidth]{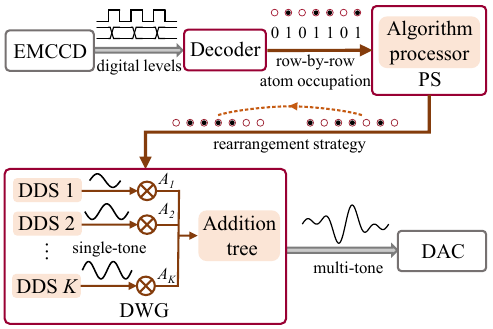}
    \caption{ 
        Design of the hardware structure. 
        The hardware modules and the peripherals are drawn in brown and grey boxes, respectively. The photon counting data for each pixel accumulated by the EMCCD is directly transferred to the on-FPGA decoder through raw logical-level signals on a row-by-row basis. The decoder determines the atom occupation and sends such information to the PS as digital signals. The PS makes the strategy based on the Tetris algorithm and informs the DWG, which then uses a set of DDSs and an addition tree to generate a multi-tone waveform. The DAC outputs RF signals to control the AODs.
    }\label{fig:processing_unit}
\end{figure}

The information flow in the hardware can be described as follows (see Fig.~\ref{fig:processing_unit}). When an atom fluorescence image is taken, the EMCCD directly sends the unprocessed pixel-wise charge accumulation data to the FPGA. Though the full-image readout speed of EMCCD is slow, it features a row-by-row charge transfer. 
The image decoder takes advantage of this readout mode and converts the logical levels received from EMCCD to atom occupation data recognizable by the PS, as soon as the image data of a single row of atom arrives.

The on-chip PS plays the role of strategy maker. It implements the Tetris algorithm and determines a rearrangement strategy once the occupation information of one row of atoms has been read out. After the strategy is made, the PS instructs the DWG module how to move the tweezers. 

The DWG then calculates the multi-tone waveform needed by AODs to generate the desired moving tweezers. The output digital waveform $S(t_n)$ for $K$ simultaneous mobile tweezers can be written as
\begin{equation}\label{eq:DWG}
     S(t_n) = \sum_{i=1}^K A_i(t_n) \cos \left(  \sum_{m=1}^{n} f_i(t_m)\Delta t+ \varphi_{0,i}\right),
\end{equation}
where $t_n=n\Delta t$ with $\Delta t$ as the clock period of DAC, $A_i$, $f_i$, and $\varphi_{0,i}$ are the amplitude, frequency, and the initial phase of the signal driving the $i$-th mobile tweezer, respectively.
Each mobile tweezer is assigned with a DDS, and each DDS consists of 8 interleaved DDS cores to fully exploit the speed of the DACs~\cite{young2020half}. This structure intrinsically guarantees that the frequencies are swept rapidly and phase-continuously. 
Therefore, the tweezers can be moved quickly and smoothly so that the atom loss during rearrangement can be minimized.
All the DDSs work in parallel and their output single-tone waveforms are rescaled and summed up in an addition-tree. 

The amplitude $A_i(t_n)$ is time-dependent for ramping up and down the tweezer intensities smoothly (Fig.~\ref{fig:principal_of_operation}(d)). We also carefully design $A_i(t_n)$ to compensate the non-uniformity of the frequency responses of AODs.

Such architecture does not require additional waveform storage, and allows almost arbitrarily many mobile tweezers once the logic unit resource requirements are met. 
As mentioned in Sec.~\ref{subsection:parallel_tweezer}, over $400$ atoms could be moved simultaneously in a row using an on-the-shelf FPGA, which is more than enough in current rearrangement experiments.

\begin{table}[t]
    \begin{ruledtabular}
    \begin{tabular}{cc}
        \textrm{Term} &
        \textrm{Time} ($\rm \mu s$) \\ \hline
        $\tau_{\text{cam}}$ & 835(5) \\
        $\tau_{\text{decoder}}$ & $37.1(1) \times N_p $ \\
        $\tau_{\text{alg}}$ & 60(3) \\
        $\tau_{\text{DWG}}$ & 0.715(5) \\
        $\tau_{\text{DAC}}$ & 0.190(5) \\
    \end{tabular}
    \end{ruledtabular}
    \caption{
        Measured starting overhead of the system for an array with 32 atoms per row. See Fig.~\ref{fig:principal_of_operation}(a) for the definition of each term. 
        Here one row of atoms is imaged onto $N_p$ rows of pixels.
    }\label{tab:latency}
\end{table}

In our characterization, we use the Camera Link output provided by our EMCCD Andor iXon 888 to directly access the charge accumulation data. The readout latency after exposure is $835(5)~\rm \mu s$. The data transfer time is $37.1(1)~\rm \mu s$ per row of pixels compared to $\sim 20~\rm ms$ for the full image readout. We choose XCKU040 from Xilinx as our FPGA platform. The PS is based on the MicroBlaze softcore. 
We select a high speed 16-bit, 4-channel DAC (DAC39J84 from Texas Instruments) and a two-dimensional AOD (DTSXY-400-850 from AA Opto-Electronic). 
Two of the DAC channels with data input rate of $1.2288\times 10^9 ~\rm samples/s$ are used to drive horizontal and vertical AODs. 
The frequency resolution is $73.24~\rm Hz$, and the minimal time step of frequency change is $6.51~\rm ns$. The measured latency between modules and the time cost of each module are summarized in Table~\ref{tab:latency}.

We experimentally demonstrate the overall performance of our system with Tetris algorithm by FPGA-in-the-loop simulations, where all steps except for atom capture and imaging are carried out on the real FPGA hardware. 
With both $\tau_{\rm cam}$ and $\tau_{\rm decoder}$ experimentally measured on our EMCCD, which are very stable and highly repeatable, we do not interface with the EMCCD in these simulations, but feed the FPGA with randomly generated initial loading configurations with a loading efficiency of $50\%$~\cite{schlosser2001sub}. Note that the FPGA still waits for $\tau_{\rm cam}$ at the beginning and for $\tau_{\rm decoder}$ before processing each row of atoms, to pretend there were real pixel data coming in, so that the total time cost reflects the actual performance of the system.
We assume each atom occupies a region of $3\times 3$ pixels in the image.
The time cost of atom moving between adjacent sites is set to $30~\rm \mu s$, and the time cost of atom transferring between static and mobile tweezers to $35~\rm \mu s$~\cite{tian2022parallel,21-256-Lukin,bluvstein_quantum_2022}. Though the total time cost is dependent on this choice, the scaling is insensitive to the atom moving speed.

\begin{video}
    \centering
    \includegraphics[width=\linewidth]{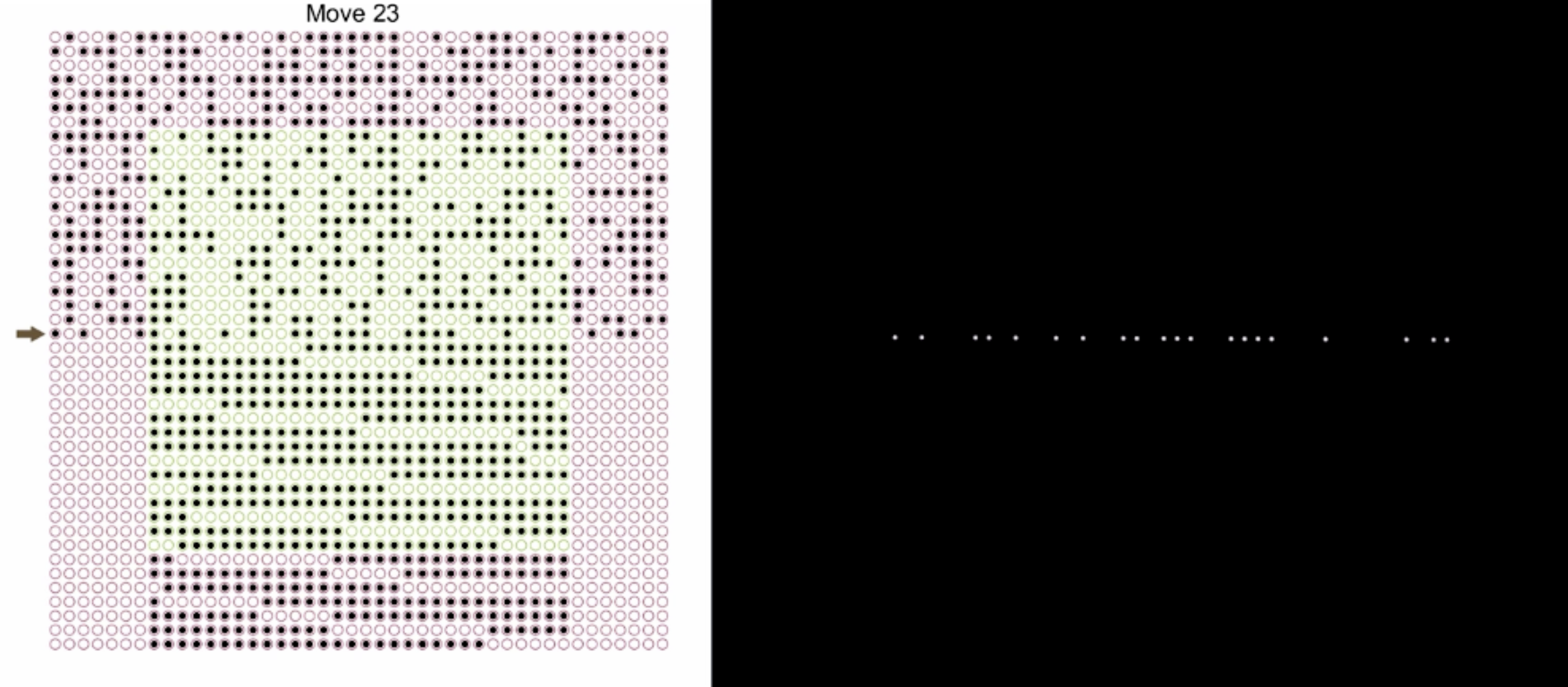}
    \setfloatlink{https://journals.aps.org/prapplied/abstract/10.1103/PhysRevApplied.19.054032}
    \caption{Example of the rearrangement of $30\times 30$ compact target geometry with a $44\times 44$ reservoir array. Left: Movement strategy given by the Tetris algorithm. Right: Image of mobile tweezers generated by the AODs.}\label{video:rearragement-example}
\end{video}

\begin{figure}[htbp]
    \centering
    \includegraphics[width=\linewidth]{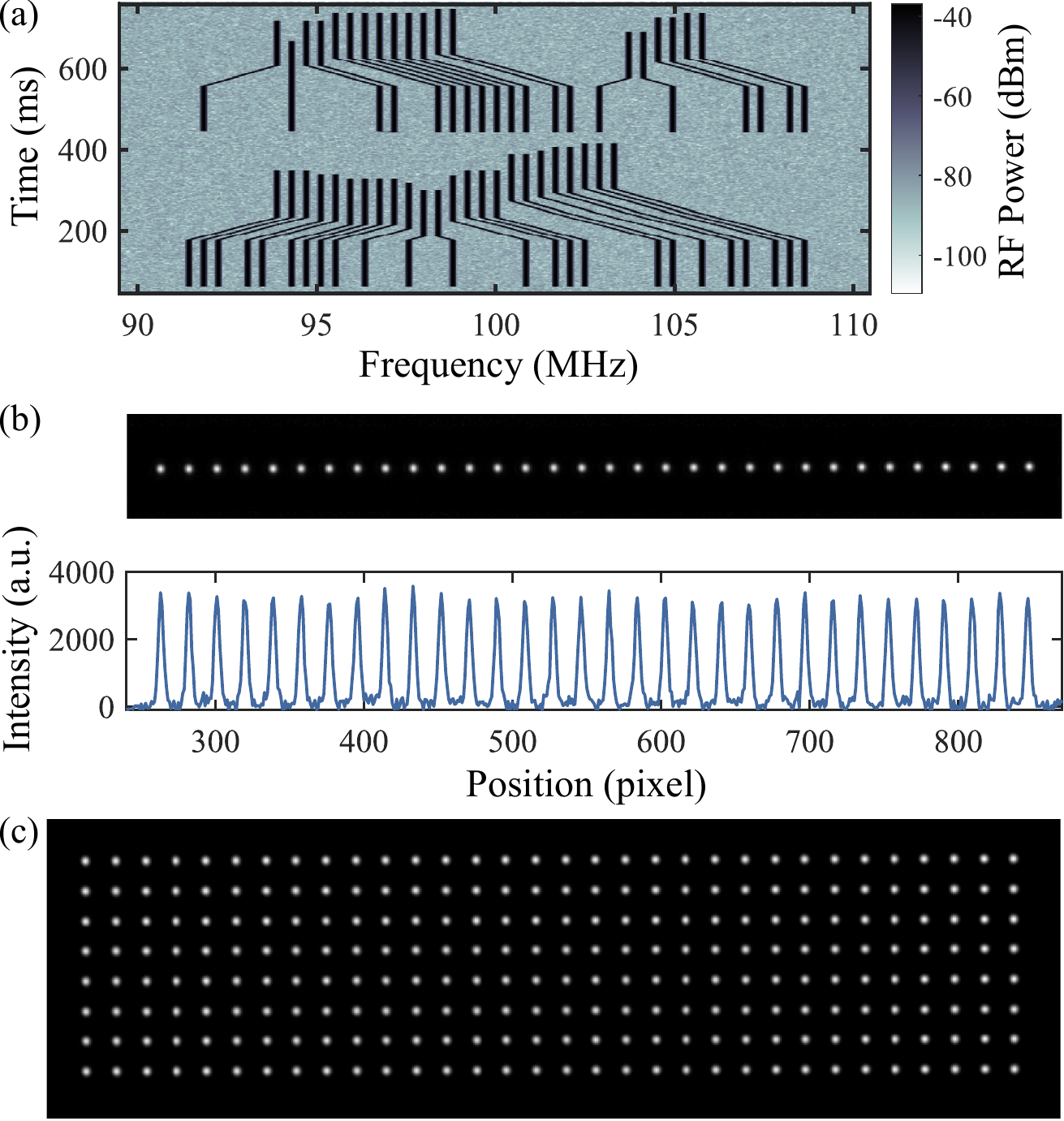}
    \caption{ 
        Experimental snapshots of the mobile tweezer generation.
        (a) The frequency spectra of the X-AOD driving signal for generating the mobile tweezers for the first two moves of the example rearrangement in the main text. The resolution bandwidth is set to $50~\rm kHz$ and the moving speed is slowed down by 30 times for better visual effect.
        (b) Snapshot of 32 AOD tweezers. The lower panel shows the integration of the intensity along the vertical direction.
        (c) Snapshot of a $32\times 8$ array of AOD tweezers. The intensity inhomogeneity in both (b) and (c) is suppressed to within $4\%$.
    }\label{fig:experiment-shots}
\end{figure}

For each number of atoms $N$ in the target array with either compact or staggered geometry, we randomly generate 500 initial atomic occupation configurations. 
Then the FPGA-based control system produces the corresponding mobile tweezers for the rearrangement and we record the total time cost.
As an example, a video demonstration for the rearrangement of $30\times 30$ compact target geometry with a $44\times 44$ reservoir array is shown in Video~\ref{video:rearragement-example}.
In Fig.~\ref{fig:experiment-shots}(a), we show the frequency spectra of the driving signals for the X-AOD to generate the corresponding mobile tweezers for the first two moves. For better visual effect, the resolution bandwidth is set to a large value of $50~\rm kHz$ and the moving speed is slowed down by 30 times. A snapshot of 32 AOD tweezers is shown in Fig.~\ref{fig:experiment-shots}(b) to demonstrate the intensity homogeneity. A larger two-dimensional $32\times 8$ tweezer array is also achievable as shown in Fig.~\ref{fig:experiment-shots}(c), though only one dimensional mobile tweezers are used in the Tetris algorithm. By randomizing initial phases of the AOD driving signals to reduce the intermodulation effect \cite{zhang_optical_2022} and compensating for the frequency dependence of AOD diffraction efficiency, the tweezer intensity variance across the entire moving region is optimized to within $4\%$ (relative standard deviation). 

\begin{figure}[htbp]
    \centering
    \includegraphics[width=\linewidth]{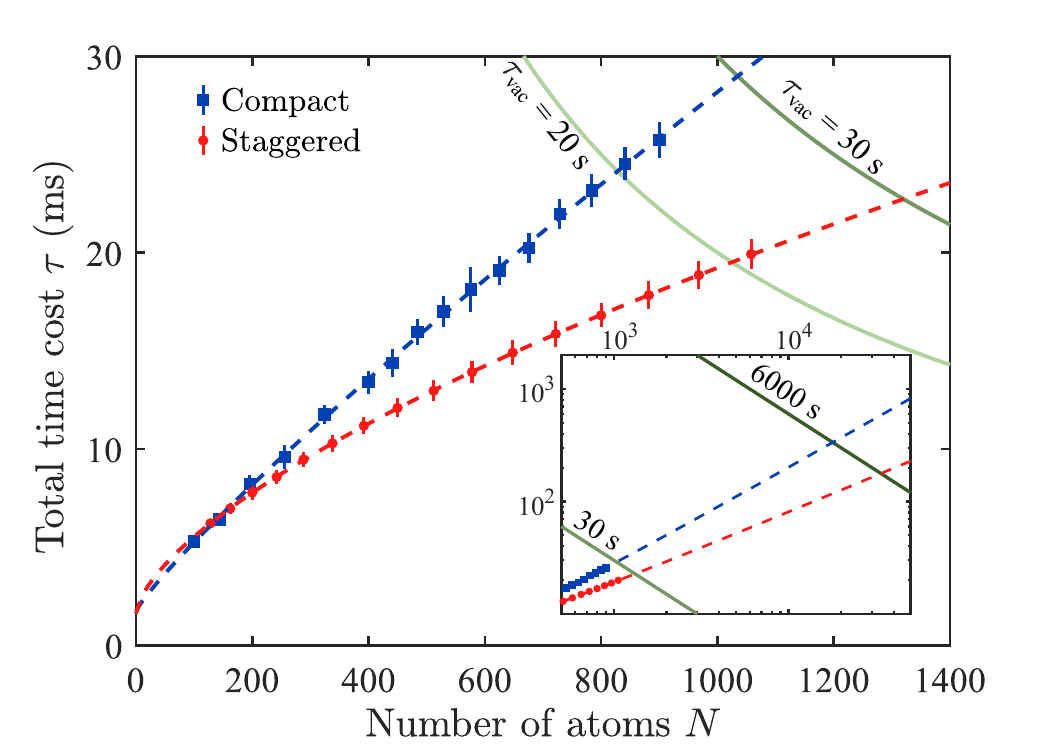}
    \caption{ 
        Experimental demonstration of the overall scalability of the rearrangement system.
        The total time costs $\tau$ as a function of $N$ to assemble a compact (blue) or staggered (red) target array are shown. Each point is averaged over 500 realizations with the error bars showing the standard deviation. The dashed lines are fittings to power functions with scaling factor of $N^{0.88(6)}$ for compact geometry, and $N^{0.65(1)}$ for staggered geometry. Note that the total time cost is depends on the moving speed of atoms, but the scaling is independent on this choice. The green lines show the collective lifetime of $N$ atoms with individual vacuum limited lifetime $30~\rm s$ (dark green) and $20~\rm s$ (light green). The intersecting points of the collective lifetime curves and the rearrangement time curves mark the lifetime-limited array size. Inset: The scalability of the rearrangement system in a broader range. The green lines correspond to individual vacuum limited atomic lifetime $6000~\rm s$ (darkest) and $30~\rm s$.
    }\label{fig:overall_performance}
\end{figure}

The overall time costs in rearrangements for compact and staggered geometries are shown in Fig.~\ref{fig:overall_performance}. The power-law fittings give a scaling factor of $N^{0.88(6)}$ for compact geometry, and $N^{0.65(1)}$ for staggered geometry. 
Compared to the algorithm in Ref.~\cite{tian2022parallel}, our algorithm has a similar scaling, though it is only based on the occupation information of the current and previous row of atoms, so that the calculation and rearrangement can start before the full image has been transferred.
The green lines show the vacuum limited ensemble lifetimes $\tau_{\text{vac}}/N$ for different single-atom vacuum limited lifetimes $\tau_{\text{vac}}$. The intersection of the two sets of curves marks the maximum number of atoms with which a defect-free geometry can be realized. 

The strategy of parallel processing during image transfer effectively reduces the total time cost and is highly scalable.
Assuming a single-atom vacuum limited lifetime in tweezer to be $\tau_{\text{vac}} = 20~\rm s$, the atom number in the defect-free atomic array would be about 830 for compact geometry and 1020 for staggered geometry.
With $\tau_{\text{vac}} = 30~\rm s$, such atom number can exceed 1000 for both geometries.

\section{conclusion and outlook}\label{sec:conclusion}
 
We enrich the quantum toolbox of large-scale defect-free atomic arrays by fully parallelizing the rearrangement process. The combination of the FPGA-based integrated hardware structure and the Tetris algorithm enables simultaneous atomic movements before the full atomic image is read out, so that the rearrangement time cost is significantly reduced. 
Given a reasonable single-atom vacuum limited lifetime of tens of seconds, defect-free atomic arrays with more than 1000 atoms are experimentally achievable. Combined with other state-of-the-art techniques, such as grey molasses~\cite{PhysRevLett.115.073003,PhysRevX.9.011057,PhysRevX.12.021027} and cryogenically enhanced vacuum for long atomic lifetime~\cite{PhysRevApplied-6000s}, assembly of 10000 atoms in defect-free arrays becomes realistic, if provided with enough optical power.

Our parallel control system can be conveniently customized to integrate other modules in the feedback loop with low latency. For example, multiple AODs can be controlled simultaneously, enabling simultaneous compilation of several different quantum circuits. Fast mid-circuit measurements, and real-time quantum feedback and feedforward, are also available~\cite{graham2022multi,singh2022mid,PhysRevApplied.12.014038,PhysRevApplied.9.034011}. We expect this flexibility would induce numerous new exciting opportunities.

\begin{acknowledgements}
    We acknowledge the financial supports from National Key Research and Development Program of China (2021YFA1400904, 2021YFA0718303) and National Natural Science Foundation of China (92165203, 61975092, 11974202).
\end{acknowledgements}

\bibliography{FPGA.bib}

\begin{thebibliography}{30}%
\makeatletter
\providecommand \@ifxundefined [1]{%
 \@ifx{#1\undefined}
}%
\providecommand \@ifnum [1]{%
 \ifnum #1\expandafter \@firstoftwo
 \else \expandafter \@secondoftwo
 \fi
}%
\providecommand \@ifx [1]{%
 \ifx #1\expandafter \@firstoftwo
 \else \expandafter \@secondoftwo
 \fi
}%
\providecommand \natexlab [1]{#1}%
\providecommand \enquote  [1]{``#1''}%
\providecommand \bibnamefont  [1]{#1}%
\providecommand \bibfnamefont [1]{#1}%
\providecommand \citenamefont [1]{#1}%
\providecommand \href@noop [0]{\@secondoftwo}%
\providecommand \href [0]{\begingroup \@sanitize@url \@href}%
\providecommand \@href[1]{\@@startlink{#1}\@@href}%
\providecommand \@@href[1]{\endgroup#1\@@endlink}%
\providecommand \@sanitize@url [0]{\catcode `\\12\catcode `\$12\catcode `\&12\catcode `\#12\catcode `\^12\catcode `\_12\catcode `\%12\relax}%
\providecommand \@@startlink[1]{}%
\providecommand \@@endlink[0]{}%
\providecommand \url  [0]{\begingroup\@sanitize@url \@url }%
\providecommand \@url [1]{\endgroup\@href {#1}{\urlprefix }}%
\providecommand \urlprefix  [0]{URL }%
\providecommand \Eprint [0]{\href }%
\providecommand \doibase [0]{https://doi.org/}%
\providecommand \selectlanguage [0]{\@gobble}%
\providecommand \bibinfo  [0]{\@secondoftwo}%
\providecommand \bibfield  [0]{\@secondoftwo}%
\providecommand \translation [1]{[#1]}%
\providecommand \BibitemOpen [0]{}%
\providecommand \bibitemStop [0]{}%
\providecommand \bibitemNoStop [0]{.\EOS\space}%
\providecommand \EOS [0]{\spacefactor3000\relax}%
\providecommand \BibitemShut  [1]{\csname bibitem#1\endcsname}%
\let\auto@bib@innerbib\@empty
%</preamble>
\bibitem [{\citenamefont {Kaufman}\ and\ \citenamefont {Ni}(2021)}]{kaufman_ni_2021}%
  \BibitemOpen
  \bibfield  {author} {\bibinfo {author} {\bibfnamefont {A.~M.}\ \bibnamefont {Kaufman}}\ and\ \bibinfo {author} {\bibfnamefont {K.-K.}\ \bibnamefont {Ni}},\ }\bibfield  {title} {\bibinfo {title} {Quantum science with optical tweezer arrays of ultracold atoms and molecules},\ }\href {https://doi.org/10.1038/s41567-021-01357-2} {\bibfield  {journal} {\bibinfo  {journal} {Nat. Phys.}\ }\textbf {\bibinfo {volume} {17}},\ \bibinfo {pages} {1324} (\bibinfo {year} {2021})}\BibitemShut {NoStop}%
\bibitem [{\citenamefont {Browaeys}\ and\ \citenamefont {Lahaye}(2020)}]{browaeys2020many}%
  \BibitemOpen
  \bibfield  {author} {\bibinfo {author} {\bibfnamefont {A.}~\bibnamefont {Browaeys}}\ and\ \bibinfo {author} {\bibfnamefont {T.}~\bibnamefont {Lahaye}},\ }\bibfield  {title} {\bibinfo {title} {Many-body physics with individually controlled {Rydberg} atoms},\ }\href {https://doi.org/10.1038/s41567-019-0733-z} {\bibfield  {journal} {\bibinfo  {journal} {Nat. Phys.}\ }\textbf {\bibinfo {volume} {16}},\ \bibinfo {pages} {132} (\bibinfo {year} {2020})}\BibitemShut {NoStop}%
\bibitem [{\citenamefont {de~Léséleuc}\ \emph {et~al.}(2019)\citenamefont {de~Léséleuc}, \citenamefont {Lienhard}, \citenamefont {Scholl}, \citenamefont {Barredo}, \citenamefont {Weber}, \citenamefont {Lang}, \citenamefont {Büchler}, \citenamefont {Lahaye},\ and\ \citenamefont {Browaeys}}]{doi:10.1126/science.aav9105}%
  \BibitemOpen
  \bibfield  {author} {\bibinfo {author} {\bibfnamefont {S.}~\bibnamefont {de~Léséleuc}}, \bibinfo {author} {\bibfnamefont {V.}~\bibnamefont {Lienhard}}, \bibinfo {author} {\bibfnamefont {P.}~\bibnamefont {Scholl}}, \bibinfo {author} {\bibfnamefont {D.}~\bibnamefont {Barredo}}, \bibinfo {author} {\bibfnamefont {S.}~\bibnamefont {Weber}}, \bibinfo {author} {\bibfnamefont {N.}~\bibnamefont {Lang}}, \bibinfo {author} {\bibfnamefont {H.~P.}\ \bibnamefont {Büchler}}, \bibinfo {author} {\bibfnamefont {T.}~\bibnamefont {Lahaye}},\ and\ \bibinfo {author} {\bibfnamefont {A.}~\bibnamefont {Browaeys}},\ }\bibfield  {title} {\bibinfo {title} {Observation of a symmetry-protected topological phase of interacting bosons with {Rydberg} atoms},\ }\href {https://doi.org/10.1126/science.aav9105} {\bibfield  {journal} {\bibinfo  {journal} {Science}\ }\textbf {\bibinfo {volume} {365}},\ \bibinfo {pages} {775} (\bibinfo {year} {2019})}\BibitemShut {NoStop}%
\bibitem [{\citenamefont {Bernien}\ \emph {et~al.}(2017)\citenamefont {Bernien}, \citenamefont {Schwartz}, \citenamefont {Keesling}, \citenamefont {Levine}, \citenamefont {Omran}, \citenamefont {Pichler}, \citenamefont {Choi}, \citenamefont {Zibrov}, \citenamefont {Endres}, \citenamefont {Greiner} \emph {et~al.}}]{bernien2017probing}%
  \BibitemOpen
  \bibfield  {author} {\bibinfo {author} {\bibfnamefont {H.}~\bibnamefont {Bernien}}, \bibinfo {author} {\bibfnamefont {S.}~\bibnamefont {Schwartz}}, \bibinfo {author} {\bibfnamefont {A.}~\bibnamefont {Keesling}}, \bibinfo {author} {\bibfnamefont {H.}~\bibnamefont {Levine}}, \bibinfo {author} {\bibfnamefont {A.}~\bibnamefont {Omran}}, \bibinfo {author} {\bibfnamefont {H.}~\bibnamefont {Pichler}}, \bibinfo {author} {\bibfnamefont {S.}~\bibnamefont {Choi}}, \bibinfo {author} {\bibfnamefont {A.~S.}\ \bibnamefont {Zibrov}}, \bibinfo {author} {\bibfnamefont {M.}~\bibnamefont {Endres}}, \bibinfo {author} {\bibfnamefont {M.}~\bibnamefont {Greiner}}, \emph {et~al.},\ }\bibfield  {title} {\bibinfo {title} {Probing many-body dynamics on a 51-atom quantum simulator},\ }\href {https://doi.org/10.1038/nature24622} {\bibfield  {journal} {\bibinfo  {journal} {Nature}\ }\textbf {\bibinfo {volume} {551}},\ \bibinfo {pages} {579} (\bibinfo {year} {2017})}\BibitemShut {NoStop}%
\bibitem [{\citenamefont {Blackmore}\ \emph {et~al.}(2018)\citenamefont {Blackmore}, \citenamefont {Caldwell}, \citenamefont {Gregory}, \citenamefont {Bridge}, \citenamefont {Sawant}, \citenamefont {Aldegunde}, \citenamefont {Mur-Petit}, \citenamefont {Jaksch}, \citenamefont {Hutson}, \citenamefont {Sauer} \emph {et~al.}}]{blackmore2018ultracold}%
  \BibitemOpen
  \bibfield  {author} {\bibinfo {author} {\bibfnamefont {J.~A.}\ \bibnamefont {Blackmore}}, \bibinfo {author} {\bibfnamefont {L.}~\bibnamefont {Caldwell}}, \bibinfo {author} {\bibfnamefont {P.~D.}\ \bibnamefont {Gregory}}, \bibinfo {author} {\bibfnamefont {E.~M.}\ \bibnamefont {Bridge}}, \bibinfo {author} {\bibfnamefont {R.}~\bibnamefont {Sawant}}, \bibinfo {author} {\bibfnamefont {J.}~\bibnamefont {Aldegunde}}, \bibinfo {author} {\bibfnamefont {J.}~\bibnamefont {Mur-Petit}}, \bibinfo {author} {\bibfnamefont {D.}~\bibnamefont {Jaksch}}, \bibinfo {author} {\bibfnamefont {J.~M.}\ \bibnamefont {Hutson}}, \bibinfo {author} {\bibfnamefont {B.}~\bibnamefont {Sauer}}, \emph {et~al.},\ }\bibfield  {title} {\bibinfo {title} {Ultracold molecules for quantum simulation: rotational coherences in {CaF} and {RbCs}},\ }\href {https://doi.org/10.1088/2058-9565/aaee35} {\bibfield  {journal} {\bibinfo  {journal} {Quantum Sci. Technol.}\ }\textbf {\bibinfo {volume} {4}},\ \bibinfo {pages} {014010} (\bibinfo {year} {2018})}\BibitemShut {NoStop}%
\bibitem [{\citenamefont {Ebadi}\ \emph {et~al.}(2021)\citenamefont {Ebadi}, \citenamefont {Wang}, \citenamefont {Levine}, \citenamefont {Keesling}, \citenamefont {Semeghini}, \citenamefont {Omran}, \citenamefont {Bluvstein}, \citenamefont {Samajdar}, \citenamefont {Pichler}, \citenamefont {Ho} \emph {et~al.}}]{21-256-Lukin}%
  \BibitemOpen
  \bibfield  {author} {\bibinfo {author} {\bibfnamefont {S.}~\bibnamefont {Ebadi}}, \bibinfo {author} {\bibfnamefont {T.~T.}\ \bibnamefont {Wang}}, \bibinfo {author} {\bibfnamefont {H.}~\bibnamefont {Levine}}, \bibinfo {author} {\bibfnamefont {A.}~\bibnamefont {Keesling}}, \bibinfo {author} {\bibfnamefont {G.}~\bibnamefont {Semeghini}}, \bibinfo {author} {\bibfnamefont {A.}~\bibnamefont {Omran}}, \bibinfo {author} {\bibfnamefont {D.}~\bibnamefont {Bluvstein}}, \bibinfo {author} {\bibfnamefont {R.}~\bibnamefont {Samajdar}}, \bibinfo {author} {\bibfnamefont {H.}~\bibnamefont {Pichler}}, \bibinfo {author} {\bibfnamefont {W.~W.}\ \bibnamefont {Ho}}, \emph {et~al.},\ }\bibfield  {title} {\bibinfo {title} {Quantum phases of matter on a 256-atom programmable quantum simulator},\ }\href {https://doi.org/10.1038/s41586-021-03582-4} {\bibfield  {journal} {\bibinfo  {journal} {Nature}\ }\textbf {\bibinfo {volume} {595}},\ \bibinfo {pages} {227} (\bibinfo {year} {2021})}\BibitemShut {NoStop}%
\bibitem [{\citenamefont {Endres}\ \emph {et~al.}(2016)\citenamefont {Endres}, \citenamefont {Bernien}, \citenamefont {Keesling}, \citenamefont {Levine}, \citenamefont {Anschuetz}, \citenamefont {Krajenbrink}, \citenamefont {Senko}, \citenamefont {Vuletic}, \citenamefont {Greiner},\ and\ \citenamefont {Lukin}}]{16-51-Lukin}%
  \BibitemOpen
  \bibfield  {author} {\bibinfo {author} {\bibfnamefont {M.}~\bibnamefont {Endres}}, \bibinfo {author} {\bibfnamefont {H.}~\bibnamefont {Bernien}}, \bibinfo {author} {\bibfnamefont {A.}~\bibnamefont {Keesling}}, \bibinfo {author} {\bibfnamefont {H.}~\bibnamefont {Levine}}, \bibinfo {author} {\bibfnamefont {E.~R.}\ \bibnamefont {Anschuetz}}, \bibinfo {author} {\bibfnamefont {A.}~\bibnamefont {Krajenbrink}}, \bibinfo {author} {\bibfnamefont {C.}~\bibnamefont {Senko}}, \bibinfo {author} {\bibfnamefont {V.}~\bibnamefont {Vuletic}}, \bibinfo {author} {\bibfnamefont {M.}~\bibnamefont {Greiner}},\ and\ \bibinfo {author} {\bibfnamefont {M.~D.}\ \bibnamefont {Lukin}},\ }\bibfield  {title} {\bibinfo {title} {Atom-by-atom assembly of defect-free one-dimensional cold atom arrays},\ }\href {https://doi.org/10.1126/science.aah3752} {\bibfield  {journal} {\bibinfo  {journal} {Science}\ }\textbf {\bibinfo {volume} {354}},\ \bibinfo {pages} {1024} (\bibinfo {year} {2016})}\BibitemShut {NoStop}%
\bibitem [{\citenamefont {Schymik}\ \emph {et~al.}(2020)\citenamefont {Schymik}, \citenamefont {Lienhard}, \citenamefont {Barredo}, \citenamefont {Scholl}, \citenamefont {Williams}, \citenamefont {Browaeys},\ and\ \citenamefont {Lahaye}}]{20Enhanced-assembly}%
  \BibitemOpen
  \bibfield  {author} {\bibinfo {author} {\bibfnamefont {K.-N.}\ \bibnamefont {Schymik}}, \bibinfo {author} {\bibfnamefont {V.}~\bibnamefont {Lienhard}}, \bibinfo {author} {\bibfnamefont {D.}~\bibnamefont {Barredo}}, \bibinfo {author} {\bibfnamefont {P.}~\bibnamefont {Scholl}}, \bibinfo {author} {\bibfnamefont {H.}~\bibnamefont {Williams}}, \bibinfo {author} {\bibfnamefont {A.}~\bibnamefont {Browaeys}},\ and\ \bibinfo {author} {\bibfnamefont {T.}~\bibnamefont {Lahaye}},\ }\bibfield  {title} {\bibinfo {title} {Enhanced atom-by-atom assembly of arbitrary tweezer arrays},\ }\href {https://doi.org/10.1103/PhysRevA.102.063107} {\bibfield  {journal} {\bibinfo  {journal} {Phys. Rev. A}\ }\textbf {\bibinfo {volume} {102}},\ \bibinfo {pages} {063107} (\bibinfo {year} {2020})}\BibitemShut {NoStop}%
\bibitem [{\citenamefont {Ebadi}\ \emph {et~al.}(2022)\citenamefont {Ebadi}, \citenamefont {Keesling}, \citenamefont {Cain}, \citenamefont {Wang}, \citenamefont {Levine}, \citenamefont {Bluvstein}, \citenamefont {Semeghini}, \citenamefont {Omran}, \citenamefont {Liu}, \citenamefont {Samajdar}, \citenamefont {Luo}, \citenamefont {Nash}, \citenamefont {Gao}, \citenamefont {Barak}, \citenamefont {Farhi}, \citenamefont {Sachdev}, \citenamefont {Gemelke}, \citenamefont {Zhou}, \citenamefont {Choi}, \citenamefont {Pichler}, \citenamefont {Wang}, \citenamefont {Greiner}, \citenamefont {Vuletić},\ and\ \citenamefont {Lukin}}]{22-289-Lukin}%
  \BibitemOpen
  \bibfield  {author} {\bibinfo {author} {\bibfnamefont {S.}~\bibnamefont {Ebadi}}, \bibinfo {author} {\bibfnamefont {A.}~\bibnamefont {Keesling}}, \bibinfo {author} {\bibfnamefont {M.}~\bibnamefont {Cain}}, \bibinfo {author} {\bibfnamefont {T.~T.}\ \bibnamefont {Wang}}, \bibinfo {author} {\bibfnamefont {H.}~\bibnamefont {Levine}}, \bibinfo {author} {\bibfnamefont {D.}~\bibnamefont {Bluvstein}}, \bibinfo {author} {\bibfnamefont {G.}~\bibnamefont {Semeghini}}, \bibinfo {author} {\bibfnamefont {A.}~\bibnamefont {Omran}}, \bibinfo {author} {\bibfnamefont {J.-G.}\ \bibnamefont {Liu}}, \bibinfo {author} {\bibfnamefont {R.}~\bibnamefont {Samajdar}}, \bibinfo {author} {\bibfnamefont {X.-Z.}\ \bibnamefont {Luo}}, \bibinfo {author} {\bibfnamefont {B.}~\bibnamefont {Nash}}, \bibinfo {author} {\bibfnamefont {X.}~\bibnamefont {Gao}}, \bibinfo {author} {\bibfnamefont {B.}~\bibnamefont {Barak}}, \bibinfo {author} {\bibfnamefont {E.}~\bibnamefont {Farhi}}, \bibinfo {author} {\bibfnamefont {S.}~\bibnamefont {Sachdev}}, \bibinfo {author} {\bibfnamefont {N.}~\bibnamefont {Gemelke}}, \bibinfo {author} {\bibfnamefont {L.}~\bibnamefont {Zhou}}, \bibinfo {author} {\bibfnamefont {S.}~\bibnamefont {Choi}}, \bibinfo {author} {\bibfnamefont {H.}~\bibnamefont {Pichler}}, \bibinfo {author} {\bibfnamefont {S.-T.}\ \bibnamefont {Wang}}, \bibinfo {author} {\bibfnamefont {M.}~\bibnamefont {Greiner}}, \bibinfo {author} {\bibfnamefont {V.}~\bibnamefont {Vuletić}},\ and\ \bibinfo {author} {\bibfnamefont {M.~D.}\ \bibnamefont {Lukin}},\ }\bibfield  {title} {\bibinfo {title} {Quantum optimization of maximum independent set using {Rydberg} atom arrays},\ }\href {https://doi.org/10.1126/science.abo6587} {\bibfield  {journal} {\bibinfo  {journal} {Science}\ }\textbf {\bibinfo {volume} {376}},\ \bibinfo {pages} {1209} (\bibinfo {year} {2022})}\BibitemShut {NoStop}%
\bibitem [{\citenamefont {Graham}\ \emph {et~al.}(2022)\citenamefont {Graham}, \citenamefont {Song}, \citenamefont {Scott}, \citenamefont {Poole}, \citenamefont {Phuttitarn}, \citenamefont {Jooya}, \citenamefont {Eichler}, \citenamefont {Jiang}, \citenamefont {Marra}, \citenamefont {Grinkemeyer} \emph {et~al.}}]{graham2022multi}%
  \BibitemOpen
  \bibfield  {author} {\bibinfo {author} {\bibfnamefont {T.}~\bibnamefont {Graham}}, \bibinfo {author} {\bibfnamefont {Y.}~\bibnamefont {Song}}, \bibinfo {author} {\bibfnamefont {J.}~\bibnamefont {Scott}}, \bibinfo {author} {\bibfnamefont {C.}~\bibnamefont {Poole}}, \bibinfo {author} {\bibfnamefont {L.}~\bibnamefont {Phuttitarn}}, \bibinfo {author} {\bibfnamefont {K.}~\bibnamefont {Jooya}}, \bibinfo {author} {\bibfnamefont {P.}~\bibnamefont {Eichler}}, \bibinfo {author} {\bibfnamefont {X.}~\bibnamefont {Jiang}}, \bibinfo {author} {\bibfnamefont {A.}~\bibnamefont {Marra}}, \bibinfo {author} {\bibfnamefont {B.}~\bibnamefont {Grinkemeyer}}, \emph {et~al.},\ }\bibfield  {title} {\bibinfo {title} {Multi-qubit entanglement and algorithms on a neutral-atom quantum computer},\ }\href {https://doi.org/10.1038/s41586-022-04603-6} {\bibfield  {journal} {\bibinfo  {journal} {Nature}\ }\textbf {\bibinfo {volume} {604}},\ \bibinfo {pages} {457} (\bibinfo {year} {2022})}\BibitemShut {NoStop}%
\bibitem [{\citenamefont {Singh}\ \emph {et~al.}(2022)\citenamefont {Singh}, \citenamefont {Bradley}, \citenamefont {Anand}, \citenamefont {Ramesh}, \citenamefont {White},\ and\ \citenamefont {Bernien}}]{singh2022mid}%
  \BibitemOpen
  \bibfield  {author} {\bibinfo {author} {\bibfnamefont {K.}~\bibnamefont {Singh}}, \bibinfo {author} {\bibfnamefont {C.~E.}\ \bibnamefont {Bradley}}, \bibinfo {author} {\bibfnamefont {S.}~\bibnamefont {Anand}}, \bibinfo {author} {\bibfnamefont {V.}~\bibnamefont {Ramesh}}, \bibinfo {author} {\bibfnamefont {R.}~\bibnamefont {White}},\ and\ \bibinfo {author} {\bibfnamefont {H.}~\bibnamefont {Bernien}},\ }\bibfield  {title} {\bibinfo {title} {Mid-circuit correction of correlated phase errors using an array of spectator qubits},\ }\Eprint {https://arxiv.org/abs/2208.11716} {arXiv:2208.11716}  (\bibinfo {year} {2022})\BibitemShut {NoStop}%
\bibitem [{\citenamefont {Kasper}\ \emph {et~al.}(2021)\citenamefont {Kasper}, \citenamefont {Gonz{\'a}lez-Cuadra}, \citenamefont {Hegde}, \citenamefont {Xia}, \citenamefont {Dauphin}, \citenamefont {Huber}, \citenamefont {Tiemann}, \citenamefont {Lewenstein}, \citenamefont {Jendrzejewski},\ and\ \citenamefont {Hauke}}]{kasper2021universal}%
  \BibitemOpen
  \bibfield  {author} {\bibinfo {author} {\bibfnamefont {V.}~\bibnamefont {Kasper}}, \bibinfo {author} {\bibfnamefont {D.}~\bibnamefont {Gonz{\'a}lez-Cuadra}}, \bibinfo {author} {\bibfnamefont {A.}~\bibnamefont {Hegde}}, \bibinfo {author} {\bibfnamefont {A.}~\bibnamefont {Xia}}, \bibinfo {author} {\bibfnamefont {A.}~\bibnamefont {Dauphin}}, \bibinfo {author} {\bibfnamefont {F.}~\bibnamefont {Huber}}, \bibinfo {author} {\bibfnamefont {E.}~\bibnamefont {Tiemann}}, \bibinfo {author} {\bibfnamefont {M.}~\bibnamefont {Lewenstein}}, \bibinfo {author} {\bibfnamefont {F.}~\bibnamefont {Jendrzejewski}},\ and\ \bibinfo {author} {\bibfnamefont {P.}~\bibnamefont {Hauke}},\ }\bibfield  {title} {\bibinfo {title} {Universal quantum computation and quantum error correction with ultracold atomic mixtures},\ }\href {https://doi.org/10.1088/2058-9565/ac2d39} {\bibfield  {journal} {\bibinfo  {journal} {Quantum Sci. Technol.}\ }\textbf {\bibinfo {volume} {7}},\ \bibinfo {pages} {015008} (\bibinfo {year} {2021})}\BibitemShut {NoStop}%
\bibitem [{\citenamefont {Preskill}(2018)}]{preskill2018quantum}%
  \BibitemOpen
  \bibfield  {author} {\bibinfo {author} {\bibfnamefont {J.}~\bibnamefont {Preskill}},\ }\bibfield  {title} {\bibinfo {title} {Quantum computing in the {NISQ} era and beyond},\ }\href {https://doi.org/10.22331/q-2018-08-06-79} {\bibfield  {journal} {\bibinfo  {journal} {Quantum}\ }\textbf {\bibinfo {volume} {2}},\ \bibinfo {pages} {79} (\bibinfo {year} {2018})}\BibitemShut {NoStop}%
\bibitem [{\citenamefont {Wilson}\ \emph {et~al.}(2022)\citenamefont {Wilson}, \citenamefont {Saskin}, \citenamefont {Meng}, \citenamefont {Ma}, \citenamefont {Dilip}, \citenamefont {Burgers},\ and\ \citenamefont {Thompson}}]{wilson2022trapping}%
  \BibitemOpen
  \bibfield  {author} {\bibinfo {author} {\bibfnamefont {J.}~\bibnamefont {Wilson}}, \bibinfo {author} {\bibfnamefont {S.}~\bibnamefont {Saskin}}, \bibinfo {author} {\bibfnamefont {Y.}~\bibnamefont {Meng}}, \bibinfo {author} {\bibfnamefont {S.}~\bibnamefont {Ma}}, \bibinfo {author} {\bibfnamefont {R.}~\bibnamefont {Dilip}}, \bibinfo {author} {\bibfnamefont {A.}~\bibnamefont {Burgers}},\ and\ \bibinfo {author} {\bibfnamefont {J.}~\bibnamefont {Thompson}},\ }\bibfield  {title} {\bibinfo {title} {Trapping alkaline earth {Rydberg} atoms optical tweezer arrays},\ }\href {https://doi.org/10.1103/PhysRevLett.128.033201} {\bibfield  {journal} {\bibinfo  {journal} {Phys. Rev. Lett.}\ }\textbf {\bibinfo {volume} {128}},\ \bibinfo {pages} {033201} (\bibinfo {year} {2022})}\BibitemShut {NoStop}%
\bibitem [{\citenamefont {Schymik}\ \emph {et~al.}(2022)\citenamefont {Schymik}, \citenamefont {Ximenez}, \citenamefont {Bloch}, \citenamefont {Dreon}, \citenamefont {Signoles}, \citenamefont {Nogrette}, \citenamefont {Barredo}, \citenamefont {Browaeys},\ and\ \citenamefont {Lahaye}}]{2022-schymik2022situ}%
  \BibitemOpen
  \bibfield  {author} {\bibinfo {author} {\bibfnamefont {K.-N.}\ \bibnamefont {Schymik}}, \bibinfo {author} {\bibfnamefont {B.}~\bibnamefont {Ximenez}}, \bibinfo {author} {\bibfnamefont {E.}~\bibnamefont {Bloch}}, \bibinfo {author} {\bibfnamefont {D.}~\bibnamefont {Dreon}}, \bibinfo {author} {\bibfnamefont {A.}~\bibnamefont {Signoles}}, \bibinfo {author} {\bibfnamefont {F.}~\bibnamefont {Nogrette}}, \bibinfo {author} {\bibfnamefont {D.}~\bibnamefont {Barredo}}, \bibinfo {author} {\bibfnamefont {A.}~\bibnamefont {Browaeys}},\ and\ \bibinfo {author} {\bibfnamefont {T.}~\bibnamefont {Lahaye}},\ }\bibfield  {title} {\bibinfo {title} {In situ equalization of single-atom loading in large-scale optical tweezer arrays},\ }\href {https://doi.org/10.1103/PhysRevA.106.022611} {\bibfield  {journal} {\bibinfo  {journal} {Phys. Rev. A}\ }\textbf {\bibinfo {volume} {106}},\ \bibinfo {pages} {022611} (\bibinfo {year} {2022})}\BibitemShut {NoStop}%
\bibitem [{\citenamefont {Schlosser}\ \emph {et~al.}(2002)\citenamefont {Schlosser}, \citenamefont {Reymond},\ and\ \citenamefont {Grangier}}]{PhysRevLett.89.023005}%
  \BibitemOpen
  \bibfield  {author} {\bibinfo {author} {\bibfnamefont {N.}~\bibnamefont {Schlosser}}, \bibinfo {author} {\bibfnamefont {G.}~\bibnamefont {Reymond}},\ and\ \bibinfo {author} {\bibfnamefont {P.}~\bibnamefont {Grangier}},\ }\bibfield  {title} {\bibinfo {title} {Collisional blockade in microscopic optical dipole traps},\ }\href {https://doi.org/10.1103/PhysRevLett.89.023005} {\bibfield  {journal} {\bibinfo  {journal} {Phys. Rev. Lett.}\ }\textbf {\bibinfo {volume} {89}},\ \bibinfo {pages} {023005} (\bibinfo {year} {2002})}\BibitemShut {NoStop}%
\bibitem [{\citenamefont {Schlosser}\ \emph {et~al.}(2001)\citenamefont {Schlosser}, \citenamefont {Reymond}, \citenamefont {Protsenko},\ and\ \citenamefont {Grangier}}]{schlosser2001sub}%
  \BibitemOpen
  \bibfield  {author} {\bibinfo {author} {\bibfnamefont {N.}~\bibnamefont {Schlosser}}, \bibinfo {author} {\bibfnamefont {G.}~\bibnamefont {Reymond}}, \bibinfo {author} {\bibfnamefont {I.}~\bibnamefont {Protsenko}},\ and\ \bibinfo {author} {\bibfnamefont {P.}~\bibnamefont {Grangier}},\ }\bibfield  {title} {\bibinfo {title} {Sub-poissonian loading of single atoms in a microscopic dipole trap},\ }\href {https://doi.org/10.1038/35082512} {\bibfield  {journal} {\bibinfo  {journal} {Nature}\ }\textbf {\bibinfo {volume} {411}},\ \bibinfo {pages} {1024} (\bibinfo {year} {2001})}\BibitemShut {NoStop}%
\bibitem [{\citenamefont {Brown}\ \emph {et~al.}(2019)\citenamefont {Brown}, \citenamefont {Thiele}, \citenamefont {Kiehl}, \citenamefont {Hsu},\ and\ \citenamefont {Regal}}]{PhysRevX.9.011057}%
  \BibitemOpen
  \bibfield  {author} {\bibinfo {author} {\bibfnamefont {M.~O.}\ \bibnamefont {Brown}}, \bibinfo {author} {\bibfnamefont {T.}~\bibnamefont {Thiele}}, \bibinfo {author} {\bibfnamefont {C.}~\bibnamefont {Kiehl}}, \bibinfo {author} {\bibfnamefont {T.-W.}\ \bibnamefont {Hsu}},\ and\ \bibinfo {author} {\bibfnamefont {C.~A.}\ \bibnamefont {Regal}},\ }\bibfield  {title} {\bibinfo {title} {Gray-molasses optical-tweezer loading: Controlling collisions for scaling atom-array assembly},\ }\href {https://doi.org/10.1103/PhysRevX.9.011057} {\bibfield  {journal} {\bibinfo  {journal} {Phys. Rev. X}\ }\textbf {\bibinfo {volume} {9}},\ \bibinfo {pages} {011057} (\bibinfo {year} {2019})}\BibitemShut {NoStop}%
\bibitem [{\citenamefont {Schymik}\ \emph {et~al.}(2021)\citenamefont {Schymik}, \citenamefont {Pancaldi}, \citenamefont {Nogrette}, \citenamefont {Barredo}, \citenamefont {Paris}, \citenamefont {Browaeys},\ and\ \citenamefont {Lahaye}}]{PhysRevApplied-6000s}%
  \BibitemOpen
  \bibfield  {author} {\bibinfo {author} {\bibfnamefont {K.-N.}\ \bibnamefont {Schymik}}, \bibinfo {author} {\bibfnamefont {S.}~\bibnamefont {Pancaldi}}, \bibinfo {author} {\bibfnamefont {F.}~\bibnamefont {Nogrette}}, \bibinfo {author} {\bibfnamefont {D.}~\bibnamefont {Barredo}}, \bibinfo {author} {\bibfnamefont {J.}~\bibnamefont {Paris}}, \bibinfo {author} {\bibfnamefont {A.}~\bibnamefont {Browaeys}},\ and\ \bibinfo {author} {\bibfnamefont {T.}~\bibnamefont {Lahaye}},\ }\bibfield  {title} {\bibinfo {title} {Single atoms with 6000-second trapping lifetimes in optical-tweezer arrays at cryogenic temperatures},\ }\href {https://doi.org/10.1103/PhysRevApplied.16.034013} {\bibfield  {journal} {\bibinfo  {journal} {Phys. Rev. Applied}\ }\textbf {\bibinfo {volume} {16}},\ \bibinfo {pages} {034013} (\bibinfo {year} {2021})}\BibitemShut {NoStop}%
\bibitem [{\citenamefont {Ohl~de Mello}\ \emph {et~al.}(2019)\citenamefont {Ohl~de Mello}, \citenamefont {Sch\"affner}, \citenamefont {Werkmann}, \citenamefont {Preuschoff}, \citenamefont {Kohfahl}, \citenamefont {Schlosser},\ and\ \citenamefont {Birkl}}]{19MLA}%
  \BibitemOpen
  \bibfield  {author} {\bibinfo {author} {\bibfnamefont {D.}~\bibnamefont {Ohl~de Mello}}, \bibinfo {author} {\bibfnamefont {D.}~\bibnamefont {Sch\"affner}}, \bibinfo {author} {\bibfnamefont {J.}~\bibnamefont {Werkmann}}, \bibinfo {author} {\bibfnamefont {T.}~\bibnamefont {Preuschoff}}, \bibinfo {author} {\bibfnamefont {L.}~\bibnamefont {Kohfahl}}, \bibinfo {author} {\bibfnamefont {M.}~\bibnamefont {Schlosser}},\ and\ \bibinfo {author} {\bibfnamefont {G.}~\bibnamefont {Birkl}},\ }\bibfield  {title} {\bibinfo {title} {Defect-free assembly of {2D} clusters of more than 100 single-atom quantum systems},\ }\href {https://doi.org/10.1103/PhysRevLett.122.203601} {\bibfield  {journal} {\bibinfo  {journal} {Phys. Rev. Lett.}\ }\textbf {\bibinfo {volume} {122}},\ \bibinfo {pages} {203601} (\bibinfo {year} {2019})}\BibitemShut {NoStop}%
\bibitem [{\citenamefont {Sheng}\ \emph {et~al.}(2021)\citenamefont {Sheng}, \citenamefont {Hou}, \citenamefont {He}, \citenamefont {Xu}, \citenamefont {Wang}, \citenamefont {Zhuang}, \citenamefont {Li}, \citenamefont {Liu}, \citenamefont {Wang},\ and\ \citenamefont {Zhan}}]{PhysRevResearch.3.023008}%
  \BibitemOpen
  \bibfield  {author} {\bibinfo {author} {\bibfnamefont {C.}~\bibnamefont {Sheng}}, \bibinfo {author} {\bibfnamefont {J.}~\bibnamefont {Hou}}, \bibinfo {author} {\bibfnamefont {X.}~\bibnamefont {He}}, \bibinfo {author} {\bibfnamefont {P.}~\bibnamefont {Xu}}, \bibinfo {author} {\bibfnamefont {K.}~\bibnamefont {Wang}}, \bibinfo {author} {\bibfnamefont {J.}~\bibnamefont {Zhuang}}, \bibinfo {author} {\bibfnamefont {X.}~\bibnamefont {Li}}, \bibinfo {author} {\bibfnamefont {M.}~\bibnamefont {Liu}}, \bibinfo {author} {\bibfnamefont {J.}~\bibnamefont {Wang}},\ and\ \bibinfo {author} {\bibfnamefont {M.}~\bibnamefont {Zhan}},\ }\bibfield  {title} {\bibinfo {title} {Efficient preparation of two-dimensional defect-free atom arrays with near-fewest sorting-atom moves},\ }\href {https://doi.org/10.1103/PhysRevResearch.3.023008} {\bibfield  {journal} {\bibinfo  {journal} {Phys. Rev. Research}\ }\textbf {\bibinfo {volume} {3}},\ \bibinfo {pages} {023008} (\bibinfo {year} {2021})}\BibitemShut {NoStop}%
\bibitem [{\citenamefont {Barredo}\ \emph {et~al.}(2016)\citenamefont {Barredo}, \citenamefont {de~Léséleuc}, \citenamefont {Lienhard}, \citenamefont {Lahaye},\ and\ \citenamefont {Browaeys}}]{16Browaeys}%
  \BibitemOpen
  \bibfield  {author} {\bibinfo {author} {\bibfnamefont {D.}~\bibnamefont {Barredo}}, \bibinfo {author} {\bibfnamefont {S.}~\bibnamefont {de~Léséleuc}}, \bibinfo {author} {\bibfnamefont {V.}~\bibnamefont {Lienhard}}, \bibinfo {author} {\bibfnamefont {T.}~\bibnamefont {Lahaye}},\ and\ \bibinfo {author} {\bibfnamefont {A.}~\bibnamefont {Browaeys}},\ }\bibfield  {title} {\bibinfo {title} {An atom-by-atom assembler of defect-free arbitrary two-dimensional atomic arrays},\ }\href {https://doi.org/10.1126/science.aah3778} {\bibfield  {journal} {\bibinfo  {journal} {Science}\ }\textbf {\bibinfo {volume} {354}},\ \bibinfo {pages} {1021} (\bibinfo {year} {2016})}\BibitemShut {NoStop}%
\bibitem [{\citenamefont {Tian}\ \emph {et~al.}(2023)\citenamefont {Tian}, \citenamefont {Wee}, \citenamefont {Qu}, \citenamefont {Lim}, \citenamefont {Datla}, \citenamefont {Koh},\ and\ \citenamefont {Loh}}]{tian2022parallel}%
  \BibitemOpen
  \bibfield  {author} {\bibinfo {author} {\bibfnamefont {W.}~\bibnamefont {Tian}}, \bibinfo {author} {\bibfnamefont {W.~J.}\ \bibnamefont {Wee}}, \bibinfo {author} {\bibfnamefont {A.}~\bibnamefont {Qu}}, \bibinfo {author} {\bibfnamefont {B.~J.~M.}\ \bibnamefont {Lim}}, \bibinfo {author} {\bibfnamefont {P.~R.}\ \bibnamefont {Datla}}, \bibinfo {author} {\bibfnamefont {V.~P.~W.}\ \bibnamefont {Koh}},\ and\ \bibinfo {author} {\bibfnamefont {H.}~\bibnamefont {Loh}},\ }\bibfield  {title} {\bibinfo {title} {Parallel assembly of arbitrary defect-free atom arrays with a multi-tweezer algorithm},\ }\href {https://doi.org/10.1103/PhysRevApplied.19.034048} {\bibfield  {journal} {\bibinfo  {journal} {Phys. Rev. Applied}\ }\textbf {\bibinfo {volume} {19}},\ \bibinfo {pages} {034048} (\bibinfo {year} {2023})}\BibitemShut {NoStop}%
\bibitem [{\citenamefont {Young}\ \emph {et~al.}(2020)\citenamefont {Young}, \citenamefont {Eckner}, \citenamefont {Milner}, \citenamefont {Kedar}, \citenamefont {Norcia}, \citenamefont {Oelker}, \citenamefont {Schine}, \citenamefont {Ye},\ and\ \citenamefont {Kaufman}}]{young2020half}%
  \BibitemOpen
  \bibfield  {author} {\bibinfo {author} {\bibfnamefont {A.~W.}\ \bibnamefont {Young}}, \bibinfo {author} {\bibfnamefont {W.~J.}\ \bibnamefont {Eckner}}, \bibinfo {author} {\bibfnamefont {W.~R.}\ \bibnamefont {Milner}}, \bibinfo {author} {\bibfnamefont {D.}~\bibnamefont {Kedar}}, \bibinfo {author} {\bibfnamefont {M.~A.}\ \bibnamefont {Norcia}}, \bibinfo {author} {\bibfnamefont {E.}~\bibnamefont {Oelker}}, \bibinfo {author} {\bibfnamefont {N.}~\bibnamefont {Schine}}, \bibinfo {author} {\bibfnamefont {J.}~\bibnamefont {Ye}},\ and\ \bibinfo {author} {\bibfnamefont {A.~M.}\ \bibnamefont {Kaufman}},\ }\bibfield  {title} {\bibinfo {title} {Half-minute-scale atomic coherence and high relative stability in a tweezer clock},\ }\href {https://doi.org/10.1038/s41586-020-3009-y} {\bibfield  {journal} {\bibinfo  {journal} {Nature}\ }\textbf {\bibinfo {volume} {588}},\ \bibinfo {pages} {408} (\bibinfo {year} {2020})}\BibitemShut {NoStop}%
\bibitem [{\citenamefont {Bluvstein}\ \emph {et~al.}(2022)\citenamefont {Bluvstein}, \citenamefont {Levine}, \citenamefont {Semeghini}, \citenamefont {Wang}, \citenamefont {Ebadi}, \citenamefont {Kalinowski}, \citenamefont {Keesling}, \citenamefont {Maskara}, \citenamefont {Pichler}, \citenamefont {Greiner}, \citenamefont {Vuletić},\ and\ \citenamefont {Lukin}}]{bluvstein_quantum_2022}%
  \BibitemOpen
  \bibfield  {author} {\bibinfo {author} {\bibfnamefont {D.}~\bibnamefont {Bluvstein}}, \bibinfo {author} {\bibfnamefont {H.}~\bibnamefont {Levine}}, \bibinfo {author} {\bibfnamefont {G.}~\bibnamefont {Semeghini}}, \bibinfo {author} {\bibfnamefont {T.~T.}\ \bibnamefont {Wang}}, \bibinfo {author} {\bibfnamefont {S.}~\bibnamefont {Ebadi}}, \bibinfo {author} {\bibfnamefont {M.}~\bibnamefont {Kalinowski}}, \bibinfo {author} {\bibfnamefont {A.}~\bibnamefont {Keesling}}, \bibinfo {author} {\bibfnamefont {N.}~\bibnamefont {Maskara}}, \bibinfo {author} {\bibfnamefont {H.}~\bibnamefont {Pichler}}, \bibinfo {author} {\bibfnamefont {M.}~\bibnamefont {Greiner}}, \bibinfo {author} {\bibfnamefont {V.}~\bibnamefont {Vuletić}},\ and\ \bibinfo {author} {\bibfnamefont {M.~D.}\ \bibnamefont {Lukin}},\ }\bibfield  {title} {\bibinfo {title} {A quantum processor based on coherent transport of entangled atom arrays},\ }\href {https://doi.org/10.1038/s41586-022-04592-6} {\bibfield  {journal} {\bibinfo  {journal} {Nature}\ }\textbf {\bibinfo {volume} {604}},\ \bibinfo {pages} {451} (\bibinfo {year} {2022})}\BibitemShut {NoStop}%
\bibitem [{\citenamefont {Zhang}\ \emph {et~al.}()\citenamefont {Zhang}, \citenamefont {Picard}, \citenamefont {Cairncross}, \citenamefont {Wang}, \citenamefont {Yu}, \citenamefont {Fang},\ and\ \citenamefont {Ni}}]{zhang_optical_2022}%
  \BibitemOpen
  \bibfield  {author} {\bibinfo {author} {\bibfnamefont {J.~T.}\ \bibnamefont {Zhang}}, \bibinfo {author} {\bibfnamefont {L.~R.~B.}\ \bibnamefont {Picard}}, \bibinfo {author} {\bibfnamefont {W.~B.}\ \bibnamefont {Cairncross}}, \bibinfo {author} {\bibfnamefont {K.}~\bibnamefont {Wang}}, \bibinfo {author} {\bibfnamefont {Y.}~\bibnamefont {Yu}}, \bibinfo {author} {\bibfnamefont {F.}~\bibnamefont {Fang}},\ and\ \bibinfo {author} {\bibfnamefont {K.-K.}\ \bibnamefont {Ni}},\ }\bibfield  {title} {\bibinfo {title} {An optical tweezer array of ground-state polar molecules},\ }\href {https://doi.org/10.1088/2058-9565/ac676c} {\bibfield  {journal} {\bibinfo  {journal} {Quantum Sci. Technol.}\ }\textbf {\bibinfo {volume} {7}},\ \bibinfo {pages} {035006}}\BibitemShut {NoStop}%
\bibitem [{\citenamefont {Lester}\ \emph {et~al.}(2015)\citenamefont {Lester}, \citenamefont {Luick}, \citenamefont {Kaufman}, \citenamefont {Reynolds},\ and\ \citenamefont {Regal}}]{PhysRevLett.115.073003}%
  \BibitemOpen
  \bibfield  {author} {\bibinfo {author} {\bibfnamefont {B.~J.}\ \bibnamefont {Lester}}, \bibinfo {author} {\bibfnamefont {N.}~\bibnamefont {Luick}}, \bibinfo {author} {\bibfnamefont {A.~M.}\ \bibnamefont {Kaufman}}, \bibinfo {author} {\bibfnamefont {C.~M.}\ \bibnamefont {Reynolds}},\ and\ \bibinfo {author} {\bibfnamefont {C.~A.}\ \bibnamefont {Regal}},\ }\bibfield  {title} {\bibinfo {title} {Rapid production of uniformly filled arrays of neutral atoms},\ }\href {https://doi.org/10.1103/PhysRevLett.115.073003} {\bibfield  {journal} {\bibinfo  {journal} {Phys. Rev. Lett.}\ }\textbf {\bibinfo {volume} {115}},\ \bibinfo {pages} {073003} (\bibinfo {year} {2015})}\BibitemShut {NoStop}%
\bibitem [{\citenamefont {Jenkins}\ \emph {et~al.}(2022)\citenamefont {Jenkins}, \citenamefont {Lis}, \citenamefont {Senoo}, \citenamefont {McGrew},\ and\ \citenamefont {Kaufman}}]{PhysRevX.12.021027}%
  \BibitemOpen
  \bibfield  {author} {\bibinfo {author} {\bibfnamefont {A.}~\bibnamefont {Jenkins}}, \bibinfo {author} {\bibfnamefont {J.~W.}\ \bibnamefont {Lis}}, \bibinfo {author} {\bibfnamefont {A.}~\bibnamefont {Senoo}}, \bibinfo {author} {\bibfnamefont {W.~F.}\ \bibnamefont {McGrew}},\ and\ \bibinfo {author} {\bibfnamefont {A.~M.}\ \bibnamefont {Kaufman}},\ }\bibfield  {title} {\bibinfo {title} {Ytterbium nuclear-spin qubits in an optical tweezer array},\ }\href {https://doi.org/10.1103/PhysRevX.12.021027} {\bibfield  {journal} {\bibinfo  {journal} {Phys. Rev. X}\ }\textbf {\bibinfo {volume} {12}},\ \bibinfo {pages} {021027} (\bibinfo {year} {2022})}\BibitemShut {NoStop}%
\bibitem [{\citenamefont {Ding}\ \emph {et~al.}(2019)\citenamefont {Ding}, \citenamefont {Cui}, \citenamefont {Huang}, \citenamefont {Li}, \citenamefont {Tu},\ and\ \citenamefont {Guo}}]{PhysRevApplied.12.014038}%
  \BibitemOpen
  \bibfield  {author} {\bibinfo {author} {\bibfnamefont {Z.-H.}\ \bibnamefont {Ding}}, \bibinfo {author} {\bibfnamefont {J.-M.}\ \bibnamefont {Cui}}, \bibinfo {author} {\bibfnamefont {Y.-F.}\ \bibnamefont {Huang}}, \bibinfo {author} {\bibfnamefont {C.-F.}\ \bibnamefont {Li}}, \bibinfo {author} {\bibfnamefont {T.}~\bibnamefont {Tu}},\ and\ \bibinfo {author} {\bibfnamefont {G.-C.}\ \bibnamefont {Guo}},\ }\bibfield  {title} {\bibinfo {title} {Fast high-fidelity readout of a single trapped-ion qubit via machine-learning methods},\ }\href {https://doi.org/10.1103/PhysRevApplied.12.014038} {\bibfield  {journal} {\bibinfo  {journal} {Phys. Rev. Applied}\ }\textbf {\bibinfo {volume} {12}},\ \bibinfo {pages} {014038} (\bibinfo {year} {2019})}\BibitemShut {NoStop}%
\bibitem [{\citenamefont {Salath\'e}\ \emph {et~al.}(2018)\citenamefont {Salath\'e}, \citenamefont {Kurpiers}, \citenamefont {Karg}, \citenamefont {Lang}, \citenamefont {Andersen}, \citenamefont {Akin}, \citenamefont {Krinner}, \citenamefont {Eichler},\ and\ \citenamefont {Wallraff}}]{PhysRevApplied.9.034011}%
  \BibitemOpen
  \bibfield  {author} {\bibinfo {author} {\bibfnamefont {Y.}~\bibnamefont {Salath\'e}}, \bibinfo {author} {\bibfnamefont {P.}~\bibnamefont {Kurpiers}}, \bibinfo {author} {\bibfnamefont {T.}~\bibnamefont {Karg}}, \bibinfo {author} {\bibfnamefont {C.}~\bibnamefont {Lang}}, \bibinfo {author} {\bibfnamefont {C.~K.}\ \bibnamefont {Andersen}}, \bibinfo {author} {\bibfnamefont {A.}~\bibnamefont {Akin}}, \bibinfo {author} {\bibfnamefont {S.}~\bibnamefont {Krinner}}, \bibinfo {author} {\bibfnamefont {C.}~\bibnamefont {Eichler}},\ and\ \bibinfo {author} {\bibfnamefont {A.}~\bibnamefont {Wallraff}},\ }\bibfield  {title} {\bibinfo {title} {Low-latency digital signal processing for feedback and feedforward in quantum computing and communication},\ }\href {https://doi.org/10.1103/PhysRevApplied.9.034011} {\bibfield  {journal} {\bibinfo  {journal} {Phys. Rev. Applied}\ }\textbf {\bibinfo {volume} {9}},\ \bibinfo {pages} {034011} (\bibinfo {year} {2018})}\BibitemShut {NoStop}%
\end{thebibliography}%



\end{document}